\pdfoutput=1
\documentclass[12pt,preprint]{aastex}
\usepackage[margin=0.75in]{geometry}
\usepackage{graphicx}
\usepackage{natbib}
\usepackage{aas_macros}
\usepackage[section] {placeins}
\usepackage{subfigure}
\usepackage{float}
\usepackage{color}

\def\msun{{\rm\,M_\odot}}

\def\msun{{\rm\,M_\odot}} 
\def\lsun{{\rm\,L_\odot}}

\newcommand{\kms}{\, {\rm km\, s}^{-1}}

\newcommand{\mpc}{\, {\rm Mpc}}
\newcommand{\kpc}{\, {\rm kpc}}
\newcommand{\hmpc}{\, h^{-1} \mpc}

\newcommand{\hi}{\mbox{H\,{\scriptsize I}\ }}
\newcommand{\heii}{\mbox{He\,{\scriptsize II}\ }}

\def\h2{${\rm\,H_2}$}

\makeatletter 

\def\hi{\noindent \hangindent=2.5em}
\def\kpc{{\rm\,kpc}}

\def\mpc{{\rm\,Mpc}}

\def\kms{{\rm\,km/s}}
\def\msun{{\rm\,M_\odot}}

\def\lsun{{\rm\,L_\odot}}

\def\vol#1  {{{#1}{\rm,}\ }}

\def\hi{{H\thinspace I}\ }

\def\heii{{He\thinspace II}\ }

\newcount\refno
\newcount\rfno
\def\eq{$^{\the\refno\ }$\advance\refno by 1}
\def\ad{\advance\rfno by 1}

\def\clock{\count0=\time \divide\count0 by 60
     \count1=\count0 \multiply\count1 by -60 \advance\count1 by \time
     \number\count0:\ifnum\count1<10{0\number\count1}\else\number\count1\fi}

\def\myputfigure#1#2#3#4#5%
{\hskip0.03\textwidth\vskip#5pt%\hfill
\makebox[0pt]{\hskip#2in
\includegraphics[width=#3\textwidth]{#1}}\vskip#4pt\hfill}

\newcount\refno
\newcount\rfno
\def\eq{$^{\the\refno\ }$\advance\refno by 1}
\def\ad{\advance\rfno by 1}

\begin{document}

\title{Infrared Properties of z=7 Galaxies from Cosmological Simulations}
 
\author{
Renyue Cen$^{1}$ and  
Taysun Kimm$^{2}$ 
} 

\footnotetext[1]{Princeton University Observatory, Princeton, NJ 08544;
 cen@astro.princeton.edu}

\begin{abstract} 

Three-dimensional panchromatic dust radiative transfer calculations are performed 
on a set of 198 galaxies of stellar masses in the range $5\times 10^8-3\times10^{10}\msun$ 
from a cosmological hydrodynamic simulation (resolved at $29$h$^{-1}$pc) at $z\sim 7$.
In a companion paper (Kimm \& Cen), the stellar mass and UV luminosity functions, and UV-optical and FUV-NUV colors
are shown to be in good agreement with observations, if an SMC-type dust extinction curve is adopted. 
Here we make useful predictions, self-consistently, of the infrared properties of these $z\sim 7$ simulated galaxies that 
can be confronted with upcoming ALMA data.
Our findings are as follows.
(1) The effective radius in the restframe MIPS$70\mu$m band is in the range of $80-400$pc proper for $z=7$ galaxies with $L_{\rm FIR}=10^{11.3-12}\lsun$.
(2) The median of the peak wavelength of the far-infrared (FIR) spectral energy distribution is in the range of $45-60\mu$m, depending on the dust-to-metal ratio.
(3) For star formation rate in the range $3-100\msun$~yr$^{-1}$ the median FIR to bolometric luminosity ratio is $60-90\%$.
(4) The FIR luminosity function displays a power law in the high end with a slope of -3.1$\pm$0.4, instead of the usual exponential decline.

\end{abstract}
 

\section{Introduction}

Did stars or quasars reionize the universe?
We do not know that for sure, although there is a likelihood
that stars most likely dominate the photoionization rate over quasars at $z\ge 6$
\citep[e.g.,][]{2008FG}.
While the final transition from an opaque to transparent universe for ionizing photons
appears to occur
at $z\sim 6$, as seen by the Sloan Digital Sky Survey (SDSS) quasar absorption spectrum observations \citep[e.g.,][]{2006Fan}, 
the reionization process may be quite complex and likely has started much earlier ($z\ge 9$),
as suggested by the high Thomson optical depth measured by 
the Wilkinson Microwave Anisotropy Probe (WMAP) observations \citep[e.g.,][]{2012Hinshaw}
and the Planck Surveyor \citep[e.g.,][]{2013Planck}.
To fundamentally answer the question of how the universe was reionized 
and what reionized it
requires a satisfactory understanding of the formation of galaxies during the epoch of reionization.

There have been rapid and remarkable advances on the observational front
to address this question, with some of the recent observations penetrating well into
the reionization era at $z\sim 6-9$ \citep[e.g.,][]{2010eBouwens,2010Bunker,2010Yan, 2010Stark, 2010Labbe, 2011Wilkins, 2011Mclure, 2012Dunlop, 2012aBouwens, 2012Finkelstein, 2013Dunlop}.
They have taught us three important things about the real universe.
First, the observed faint end slope of the galaxy luminosity function (LF) is close to $-2$,
indicating that fainter galaxies below the current detection limit 
likely make a significant contribution to the overall ionizing photon budget.
Second, it appears that star formation in these systems may have started much earlier \citep[][]{2010Labbe}.
Third, the stellar mass of detected galaxies is at or above $10^{9}\msun$,
implying halo masses of $\ge 10^{10}\msun$ at the current detection limit.
The standard cosmological constant-dominated cold dark matter model (LCDM) \citep[e.g.,][]{1995Krauss} 
predicts that the majority of stars at $z=6-10$ are in small dwarf galaxies residing in halos 
of mass $\sim 10^{8}-10^{10}\msun$ \citep[e.g.,][]{2009Wise}.
Thus, it seems likely that the candidate galaxies that are currently detected in the Hubble Ultra Deep Field
and Early Release Science (ERS) observations at $z\sim 7-8$ 
may represent the high end of the galaxy mass spectrum.

Can the multi-wavelength predictions of the standard cold dark matter model reproduce the observations?
To answer this question, we have performed 
panchromatic ($\lambda=0.1-1000\mu$m) three-dimensional dust radiative transfer 
calculations on 198 galaxies of stellar mass $5\times 10^8-3\times10^{10}\msun$ 
obtained from an ab initio cosmological adaptive mesh refinement hydrodynamic simulation 
with high resolution ($29$h$^{-1}$pc). 
Three parameters are used in the radiative transfer calculation: the dust-to-metal ratio, the extinction curve, 
and the fraction of directly escaped light from stars ($f_{\rm esc}$). 
In \citet{2013Kimm} we show that our stellar mass function is in broad agreement with \citet{2011Gonzalez}, independent of the three parameters. 
We also show that our simulated galaxies can reasonably and simultaneously match the 
observed UV-optical color, UV spectral slope and the 
UV luminosity function, if $f_{\rm esc}\sim10\%$ and a Small Magellanic Cloud (SMC)-type extinction curve is used. 
The conclusion that the model with $f_{\rm esc}\sim 10\%$ is favored over the model with much smaller $f_{\rm esc}$ 
is encouraging based on independent considerations.
Observations of cosmological reionization infer the Thomson optical 
depth $\tau_e = 0.089 \pm 0.014$ \citep[][]{2012Hinshaw}, indicating a reionization
redshift of $z_{re}=10.6 + 1.1$ (assuming a sudden reionization picture). 
In order to reionize the universe at this redshift range by stellar sources, 
$f_{\rm esc} \ll 10\%$ would not be viable \citep[e.g.,][]{2003Cen}.
Moreover, detailed radiative transfer simulations with still higher resolutions indicate a porous interstellar medium, and
$f_{\rm esc}\sim 10\%$ is within the range of predictions for galaxies at high redshift
\citep[e.g.,][]{2009Wise, 2010Razoumov, 2011Yajima}.

One interesting and perhaps not entirely expected property,
according to the conventional wisdom,
that comes from our calculations is that most of these simulated galaxies are heavily dust-attenuated 
with extinction of $1\le A_{\rm FUV}\le 5$.
This indicates that a significant amount of stellar radiation may be re-processed by
dust and seen in the FIR.
The questions are then: 
Does the star formation rate (SFR) inferred
based on the observed optical and UV luminosities nearly account for the total SFR? 
What are their infrared properties that can be checked by upcoming ALMA observations
\citep[e.g.,][]{2008Carilli,2013Hodge}?
This paper addresses these two questions.
The outline of this paper is as follows.
In \S 2 we detail our simulations (\S 2.1), method of making galaxy catalogs (\S 2.2) and 
panchromatic three-dimensional dust radiative transfer method (\S 2.3).
Results are presented in \S 3.
Conclusions are given in \S 4.

\section{Simulations}\label{sec: sims}

\subsection{Hydrocode and Simulation Parameters}

The cosmological simulation is performed with the Eulerian hydrodynamics code, {\sc enzo} \citep[][]{1999aBryan, 2005OShea, 2009Joung}.
For more details on the simulation setup and implemented physics, the reader is referred to \citet[][]{2012Cen}.
We use the following cosmological parameters that are consistent with 
the WMAP7-normalized \citep[][]{2010Komatsu} $\Lambda$CDM model:
$\Omega_M=0.28$, $\Omega_b=0.046$, $\Omega_{\Lambda}=0.72$, $\sigma_8=0.82$,
$H_0=100 h \,{\rm km\, s}^{-1} {\rm Mpc}^{-1} = 70 \,{\rm km\, s}^{-1} {\rm Mpc}^{-1}$ and $n=0.96$.
These parameters are consistent with those from Planck first-year data \citep[][]{2013Planck}
if we average Planck derived $H_0$ with SN Ia and HST based $H_0$.
First we ran a low resolution simulation with a periodic box of $120~h^{-1}$Mpc 
(comoving) on a side.
We identified a region centered on a cluster of mass of $\sim 3\times 10^{14}\msun$ at $z=0$.
We then resimulate with high resolution of the chosen region embedded
in the outer $120h^{-1}$Mpc box to properly take into account the large-scale tidal field
and appropriate boundary conditions at the surface of the refined region.
The refined region %for ``C" run 
has a comoving size of $21\times 24\times 20h^{-3}$Mpc$^3$ 
and represents $1.8\sigma$ matter density fluctuation on that volume.
The dark matter particle mass in the refined region is $1.3\times 10^7h^{-1}\msun$.
The refined region is surrounded by three layers (each of $\sim 1h^{-1}$Mpc) of buffer zones with 
particle masses successively larger by a factor of $8$ for each layer, 
which then connects with
the outer root grid that has a dark matter particle mass $8^4$ times that in the refined region.
We choose the mesh refinement criterion such that the resolution is 
always better than $29$h$^{-1}$pc (physical), corresponding to a maximum mesh refinement level of $13$ at $z=0$.
The simulations include
a metagalactic UV background
\citep[][]{1996Haardt} where the cosmic reionization occurs at $z=9$,
and a model for shielding of UV radiation \citep[][]{2005Cen}.
They include metallicity-dependent radiative cooling \citep[][]{1995Cen}.
Our simulations also solve relevant gas chemistry
chains for molecular hydrogen formation \citep[][]{1997Abel},
molecular formation on dust grains \citep[][]{2009Joung},
and metal cooling extended down to $10~$K \citep[][]{1972Dalgarno}.
Star particles are created in cells that satisfy a set of criteria for 
star formation proposed by \citet[][]{1992CenOstriker}.
Each star particle is tagged with its initial mass, creation time, and metallicity; 
star particles typically have masses of $\sim$$10^6\msun$.

Supernova feedback from star formation is modeled following \citet[][]{2005Cen}.
Feedback energy and ejected metal-enriched mass are distributed into 
27 local gas cells centered at the star particle in question, 
weighted by the specific volume of each cell. 
This is to mimic the physical process of supernova
blastwave propagation that tends to channel energy, momentum and mass into the least dense regions
(with the least resistance and cooling).
The primary advantages of this supernova energy based feedback mechanism are three-fold.
First, nature does drive winds in this way and energy input is realistic.
Second, it has only one free parameter $e_{SN}$, namely, the fraction of the rest mass energy of stars formed
that is deposited as thermal energy on the cell scale at the location of supernovae.
Third, the processes are treated physically, obeying their respective conservation laws (where they apply),
allowing transport of metals, mass, energy and momentum to be treated self-consistently
and taking into account relevant heating/cooling processes at all times.
We allow the entire feedback processes to be hydrodynamically coupled to surroundings
and subject to relevant physical processes, such as cooling and heating. %, as in nature.
The total amount of explosion kinetic energy from Type II supernovae
for an amount of star formed $M_{*}$
with a Chabrier initial mass function (IMF) is $e_{SN} M_* c^2$ (where $c$ is the speed of light)
with  $e_{SN}=6.6\times 10^{-6}$.
Taking into account the contribution of prompt Type I supernovae,
we use $e_{SN}=1\times 10^{-5}$ in our simulations.
Observations of local starburst galaxies indicate
that nearly all of the star formation produced kinetic energy 
is used to power galactic superwinds \citep[e.g.,][]{2001Heckman}. 
Supernova feedback is important primarily for regulating star formation
and for transporting energy and metals into the intergalactic medium.
The extremely inhomogeneous metal enrichment process
demands that both metals and energy (and momentum) are correctly modeled so that they
are transported in a physically sound (albeit still approximate at the current resolution) way.

\subsection{Simulated Galaxy Catalogs}

We identify galaxies in our high resolution simulations using the HOP algorithm 
\citep[][]{1999Eisenstein}, operated on the stellar particles, which is tested to be robust
and insensitive to specific choices of concerned parameters within reasonable ranges.
%{\rm For the HOP program we choose an outer
%stellar overdensity of $\delta_{outer}=10000$, and density ratios of
%$\delta_{saddle}=2.5\delta_{outer}$ and 
%$\delta_{peak}=3\delta_{outer}$ \citet{Eisenstein99}.} 
Satellites within a galaxy are clearly identified separately.
The luminosity of each stellar particle at each of the Sloan Digital Sky Survey (SDSS) five bands 
is computed using the GISSEL stellar synthesis code \citep[][]{Bruzual03}, 
by supplying the formation time, metallicity and stellar mass.
Collecting luminosity and other quantities of member stellar particles, gas cells and dark matter 
particles yields
the following physical parameters for each galaxy:
position, velocity, total mass, stellar mass, gas mass, 
mean formation time, 
mean stellar metallicity, mean gas metallicity,
star formation rate,
luminosities in five SDSS bands (and various colors) and others.

\subsection{Three-Dimensiona Panchromatic Dust Radiative Transfer Calculations}

%To compute the spectral energy distributions (SEDs) of each galaxy,
%we post-process the simulation output at $z=7$ using a three-dimensional dust radiation transfer code,
%{\sc sunrise} \citep{jonsson06,jonsson10}. The main advantage of {\sc sunrise} is the use of a
%polychromatic algorithm, which can trace information in all wavelengths per each ray.
%It makes use of the standard dust cross-sections \citep[e.g.,][]{weingartner01,draine07} to
%simulate absorption and multiple scattering by dust.
%The input stellar spectrum is taken from {\sc starburst99} \citep{leitherer99}
%assuming a Kroupa initial mass function with the low (high) mass cut-off of $0.1 \msun\ (100 \msun)$.
%{\sc sunrise} also uses the spectrum of {\sc Hii} or photo-dissociation regions (PDRs) computed by a
%photo-ionization code, {\sc mappingsiii} \citep{dopita05,groves08}, to take into account the
%immediate absorption and emission by birth clouds. This is done by replacing SEDs of
%young ($\le 10\,{\rm Myr}$) star particles with re-processed SEDs of a population with
%constant star formation for 10 Myr by {\sc mappingsiii} \citep[see][]{jonsson10}.
%The fraction of light processed by PDRs is controlled by a
%parameter, $f_{\rm PDR}$, which we adopt $f_{\rm PDR}=0.2$ following \citet{jonsson10}.
%The amount of dust derived from the amount of metals in each simulation cell
%by using a dust-to-metal ratio ($D/M$).
%While the UV/optical properties depend sensitively on $D/M$, the FIR properties depends
%on $D/M$ only weakly, as will be shown, making our predictions of FIR properties of $z=7$ rather robust.

We post-process the simulated galaxy sample at $z=7$ using a three-dimensional dust radiation transfer code,
{\sc sunrise} \citep{jonsson06,jonsson10}.
The main strength of the {\sc sunrise} code is the use of a polychromatic algorithm to trace 
information in all wavelengths per ray, enabling us to compute the spectral energy distributions (SEDs) of each galaxy. 
It makes use of the standard dust cross-sections by Draine and collaborators \citep[][]{weingartner01,draine07} to
simulate absorption and multiple scattering by dust.
Each stellar particle that is basically a coeval star cluster 
has three attributes - mass ($\sim 10^{4-5}\msun$), formation time and metallicity -
which are input to the code {\sc starburst99} \citep{leitherer99},
assuming a Kroupa initial mass function with an upper (lower) mass limit of $100 \msun\ (0.1 \msun)$.
The output from {\sc starburst99} is 
the input stellar spectrum to {\sc sunrise}.
In order to take into account the immediate absorption and emission by birth clouds, 
which large-scale cosmological simulations cannot resolve, {\sc sunrise} uses the spectra of 
{\sc Hii} and photo-dissociation regions (PDRs) computed by 
photo-ionization code, {\sc mappingsiii} \citep{dopita05,groves08}.
This is done by replacing SEDs of star particles younger than $10\,{\rm Myr}$ with re-processed SEDs 
of a population with constant star formation for 10 Myr by {\sc mappingsiii} \citep[see][]{jonsson10}.
The fraction of light travelled through the PDR is controlled by a
parameter, $f_{\rm PDR}$, which we use $f_{\rm PDR}=0.2$ following \citet{jonsson10}.
The metal mass in each simulation cell is followed hydrodynamically,
including flux, sources (from stellar feedback) and sinks (forming into new stars).
The amount of dust is derived from the amount of metals in each hydro cell
by using a dust-to-metal ratio ($D/M$).
While the UV/optical properties depend sensitively on $D/M$, the FIR properties depends
on $D/M$ only weakly, as will be shown, making our predictions of FIR properties of $z=7$ rather robust,
except the peak wavelength of the FIR SED (see Figure~\ref{fig:peak} below).
The dust temperature, emission, and dust opacity is obtained
in a self-consistent fashion in SUNRISE by solving for the thermal
equilibrium solution for dust grains at every location, where dust emission cooling is balanced by 
stellar radiative heating.
This is achieved by iteration computationally.

\section{Results}

\begin{figure}[!h]    %Figure 1
\centering
\vskip 0.5cm
%\resizebox{5.0in}{!}{\includegraphics[angle=0]{galsed.pdf}} 
%\vskip 0.5cm
%\resizebox{5.0in}{!}{\includegraphics[angle=0]{galimg.pdf}} 
\hskip -1cm
\resizebox{3.4in}{!}{\includegraphics[angle=0]{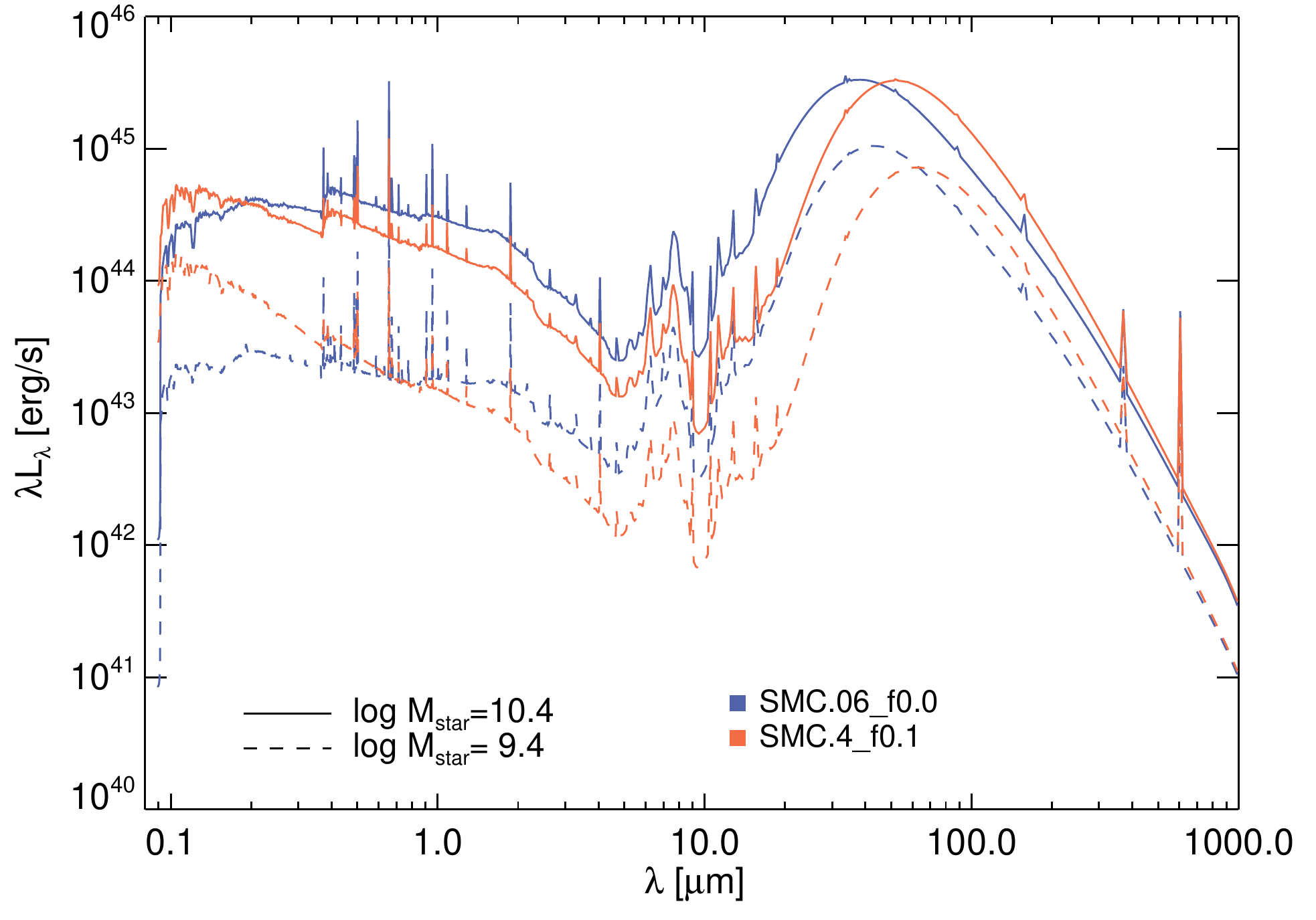}} 
\hskip -0cm
\resizebox{3.9in}{!}{\includegraphics[angle=0]{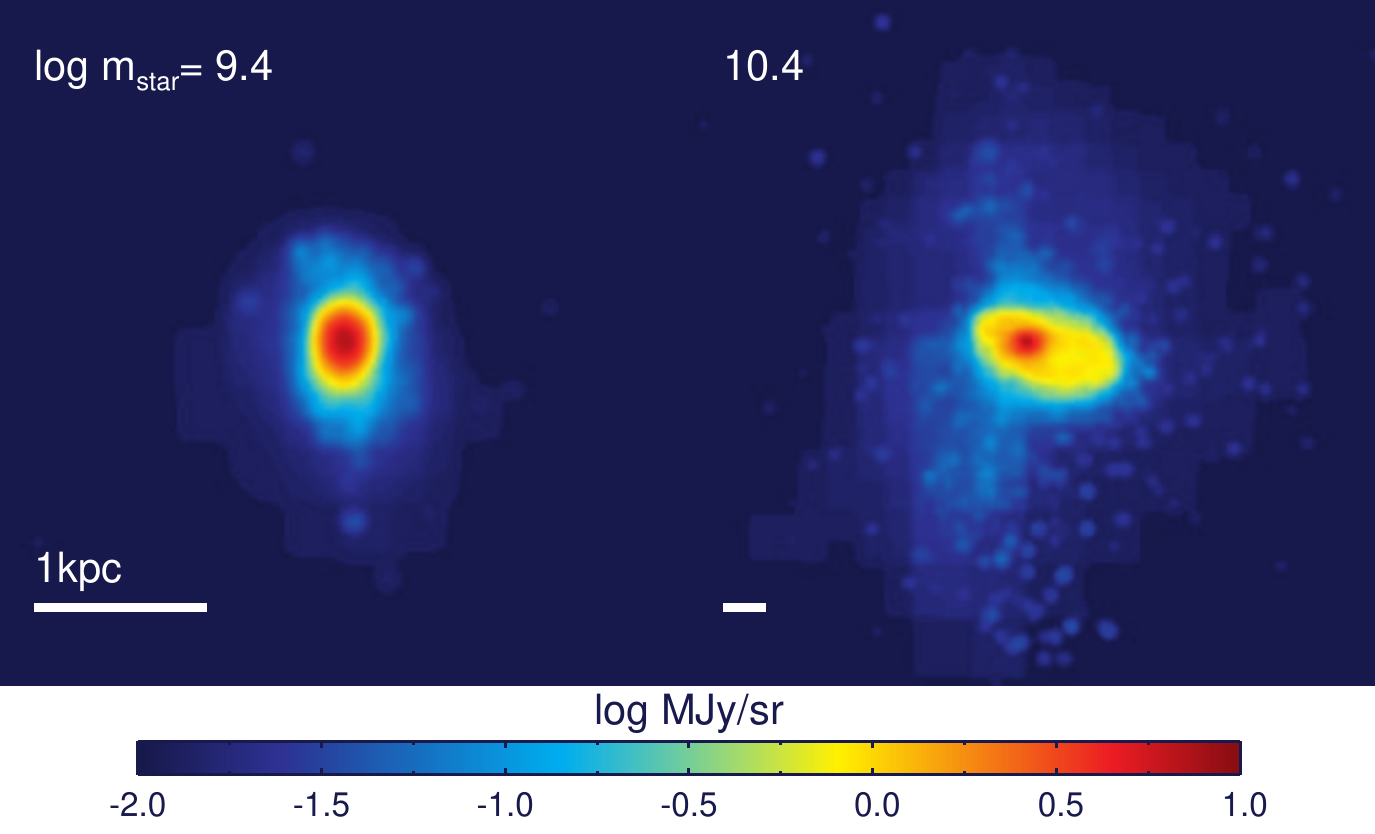}} 
\vskip -0cm
\caption{%\footnotesize %\scriptsize
Left panel: the rest-frame spectra (left panel) of two simulated galaxies of stellar masses of $10^{9.4}\msun$ (dashed curves)
and $10^{10.4}\msun$ (solid curves), respectively, at $z=7$.
The star formation rate (SFR) and specific SFR (sSFR) for the low and high mass galaxies 
are (${\rm 94\,M_{\odot}/yr}$, ${\rm 3.7\,Gyr^{-1}}$) and (${\rm 35\,M_{\odot}/yr}$, ${\rm 13\,Gyr^{-1}}$), respectively.
For each galaxy we show two cases with two different combinations of ($D/M$, $f_{\rm esc}$), (0.4,0.1) (red) and (0.06,0) (blue),
both of which are found to reproduce the UV/optical properties of $z=7$ galaxies \citep{2013Kimm}.
Right panel: restframe MIPS 70$\mu$m band images, smoothed with a Gaussian of FWHM=0.1kpc.
The number in each panel indicates logarithmic stellar mass and white ticks in both panels are 1kpc proper.
}
\label{fig:galsed}
\end{figure}

\begin{figure}[!h]    %Figure 2
\centering
\vskip -0cm
\hskip -1cm
%\resizebox{3.6in}{!}{\includegraphics[angle=0]{Peak1.pdf}} 
%\hskip -0cm
%\resizebox{3.6in}{!}{\includegraphics[angle=0]{Peak2.pdf}} 
\resizebox{3.3in}{!}{\includegraphics[angle=0]{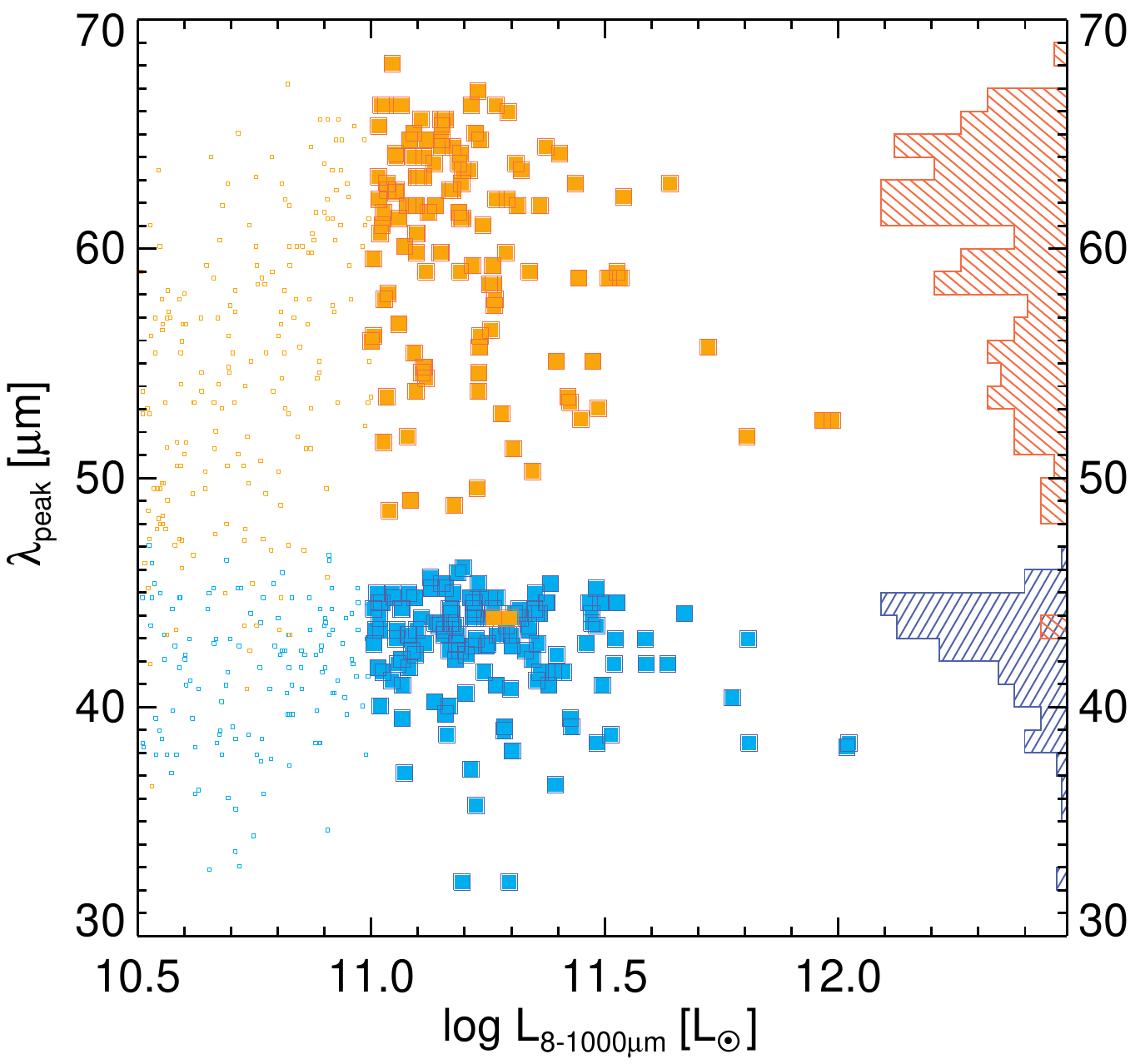}} 
\hskip -0cm
\resizebox{3.3in}{!}{\includegraphics[angle=0]{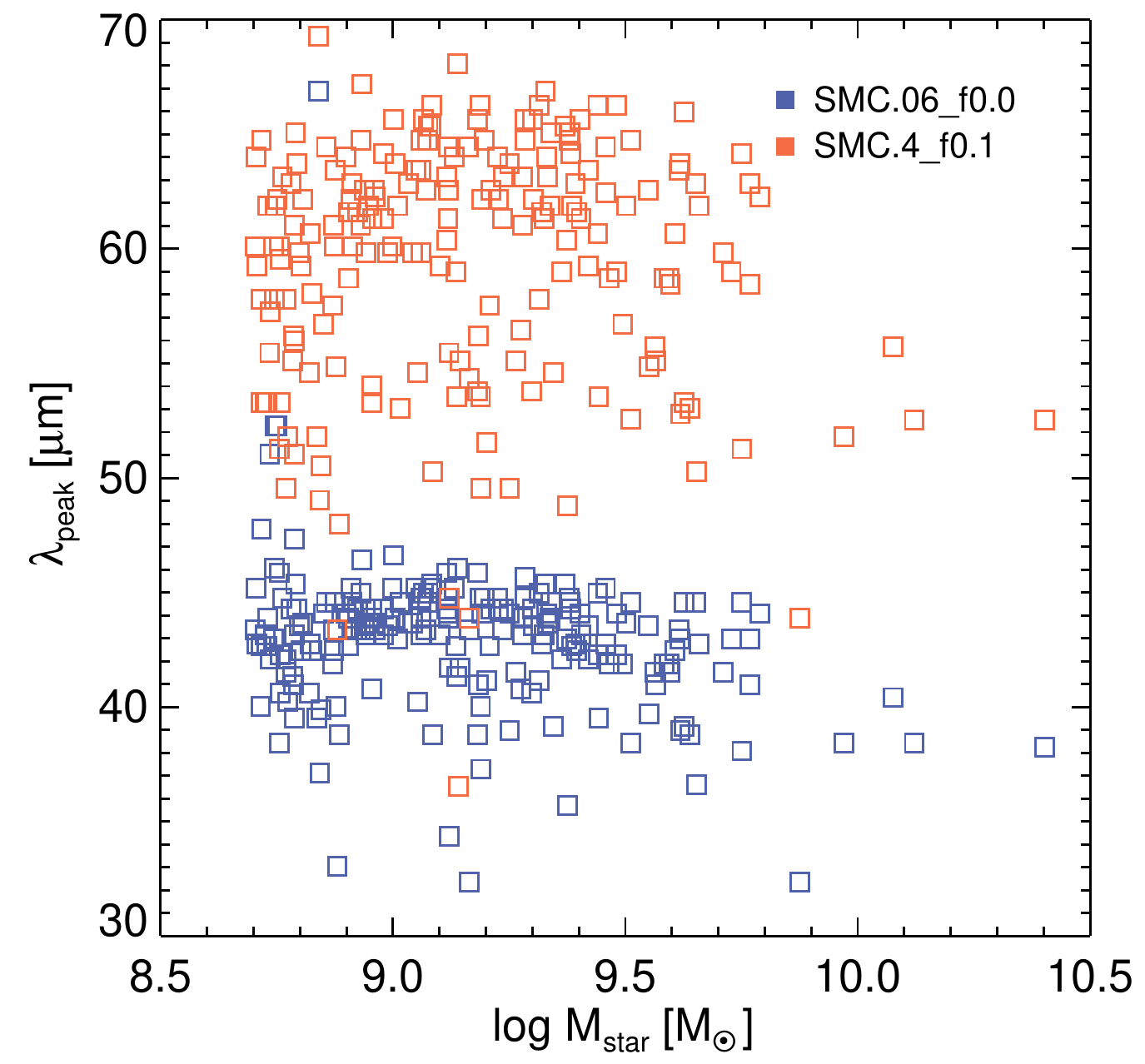}} 
\vskip -0cm
\caption{%\footnotesize %\scriptsize
shows the FIR peak wavelength $\lambda_{\rm peak}$ of the SED of galaxies at $z=7$ 
as a function of FIR luminosity (left panel) and stellar mass (right panel).
Only galaxies with stellar masses $M_{\rm star} \ge 5\times 10^8 M_{\odot}$ are shown.
For each galaxy two cases are shown with two different combinations of ($D/M$, $f_{\rm esc}$), (0.4,0.1) (red) and (0.06,0) (blue).
The distributions of $\lambda_{\rm peak}$ are shown on the right y-axis for the two cases.
The histogram in the left panel is plotted using the sample with $L_{8-1000\mu m} \ge 10^{11} L_{\odot}$
over which our sample is more than 80\% complete.
The completeness is estimated as follows. 
At IR luminosity of $\ge 10^{11} L_{\odot}$
we compute the fraction of galaxies with stellar masses
larger than $5\times 10^8\msun$ (which is our stellar mass threshold deemed reliably resolved), which is found to be 80\%. 
}
\label{fig:peak}
\end{figure}

We first present in Figure~\ref{fig:galsed} the SED in the rest-frame and 
rest-frame FIR images of two simulated galaxies of stellar masses of $10^{9.4}\msun$ and $10^{10.4}\msun$, respectively.
For each galaxy we show two cases with two different combinations of ($D/M$, $f_{\rm esc}$), (0.4,0.1) (red) and (0.06,0) (blue),
which have been both found to reproduce the UV/optical properties of $z=7$ galaxies \citep{2013Kimm}.
Three points are worth noting.
First, the overall SED shares some of the properties of lower redshift starburst galaxies,
in that the majority of the stellar radiation comes out in the FIR peak.
Second, even though the two galaxies are in the category of starburst galaxies,
the higher mass galaxy contains a significant, evolved stellar population, as indicated by a strong Balmer break at $\sim 0.37\mu$m;
the lower mass galaxy has a less prominent Balmer break, suggesting an overall younger population.
The existence of an evolved population for $z\sim 7$ galaxies is consistent with observations \citep[e.g.,][]{2010Labbe}.
It is likely that our simulation has underestimated the star formation rate at progressively higher redshift
due to poorer resolutions for smaller objects at higher redshift, hence the evolved population.
Third, the wavelength $\lambda_{\rm peak}$ of the FIR peak at $45-65\mu$m suggests a relatively hot emitting dust,
due to a combination of relatively high SFR and sub-kpc, compact sizes of the starbursting regions,
seen in the images in the right panel.

Figure~\ref{fig:peak} shows the FIR peak wavelength $\lambda_{\rm peak}$ of the SED of the galaxies at $z=7$ 
as a function of FIR luminosity (left panel) and stellar mass (right panel).
For each galaxy two cases are shown with two different combinations of ($D/M$, $f_{\rm esc}$), (0.4,0.1) (red) and (0.06,0) (blue).
The distributions of $\lambda_{\rm peak}$ are shown on the right y-axis for the two cases.
Two points are noted.
First, at the lower stellar mass ($\le 10^{10}\msun$) or lower FIR luminosity end ($L_{\rm FIR}\le 10^{11.5}\lsun$),
no strong correlation is found between either FIR luminosity or stellar mass and $\lambda_{\rm peak}$,
whereas there is some hint that the highest FIR luminosity 
($L_{\rm FIR}\ge 10^{11.5}\lsun$)
or highest stellar mass ($\ge 10^{10}\msun$) 
galaxies tend to occupy the low end of the $\lambda_{\rm peak}$ distribution.
Second, for a given simulated galaxy, post-processing it with different dust-to-metal ratios $D/M$
yields significant difference in the $\lambda_{\rm peak}$ distributions.
For the lower $D/M=0.06$ case, the median of the $\lambda_{\rm peak}$ distribution is $43\mu$m,
compared to $61\mu$m for the case with $D/M=0.4$.
This significant dependence of $\lambda_{\rm peak}$ on $D/M$ may provide a probe
of the latter, when the former is observationally obtained.
The physical origin for this, in simple terms, is that a lower $D/M$ value gives a lower dust opacity
hence a higher interstellar radiation field, which in turn heats up the dust to a higher
equilibrium temperature (balanced by emission cooling, primarily).

\begin{figure}[!h]   %Figure 3
\centering
\vskip -0.0cm
\hskip -1cm
%\resizebox{3.6in}{!}{\includegraphics[angle=0]{Ruv.pdf}} 
%\hskip -0cm
%\resizebox{3.6in}{!}{\includegraphics[angle=0]{Rir.pdf}} 
\resizebox{3.3in}{!}{\includegraphics[angle=0]{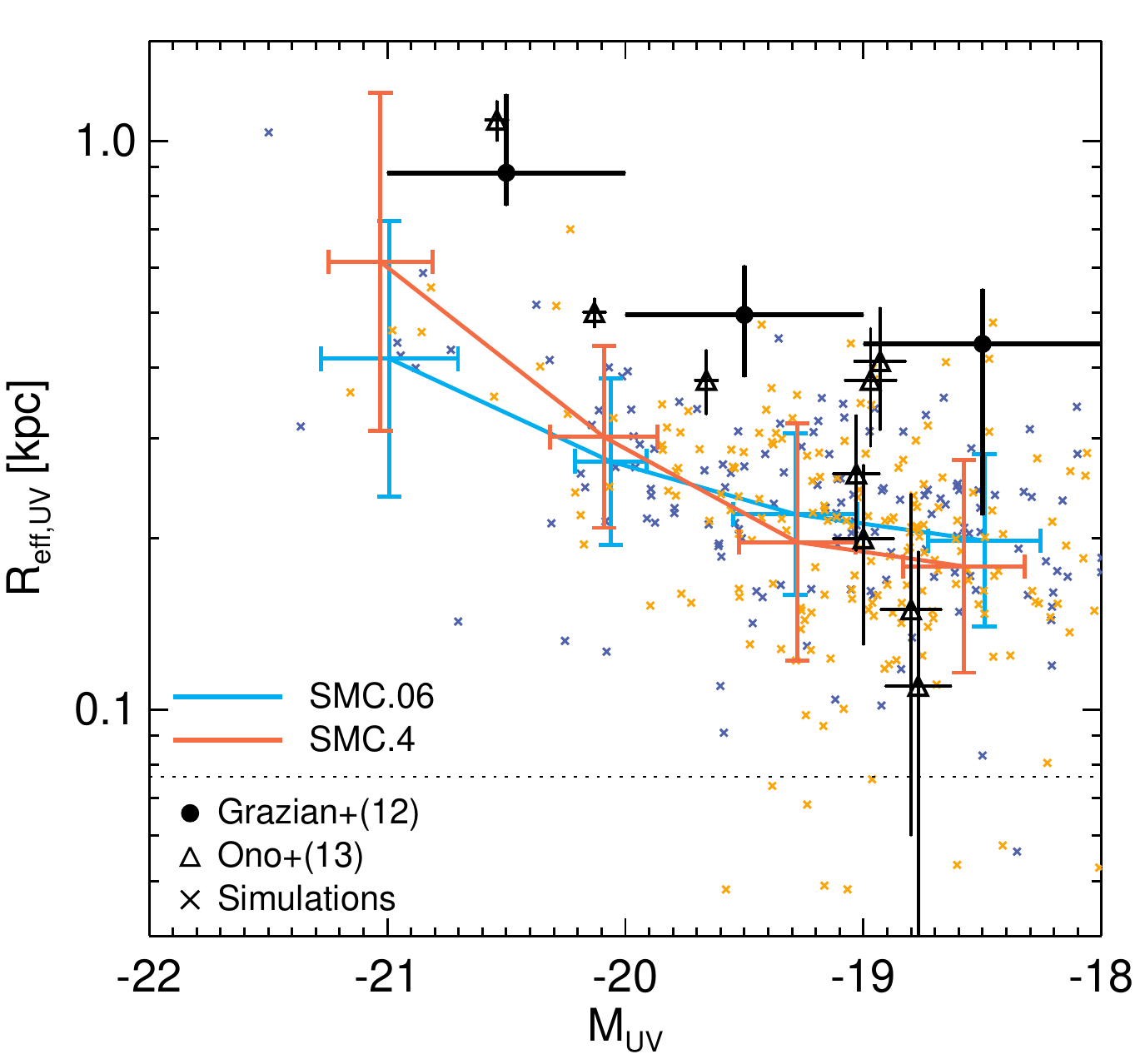}} 
\hskip -0cm
\resizebox{3.3in}{!}{\includegraphics[angle=0]{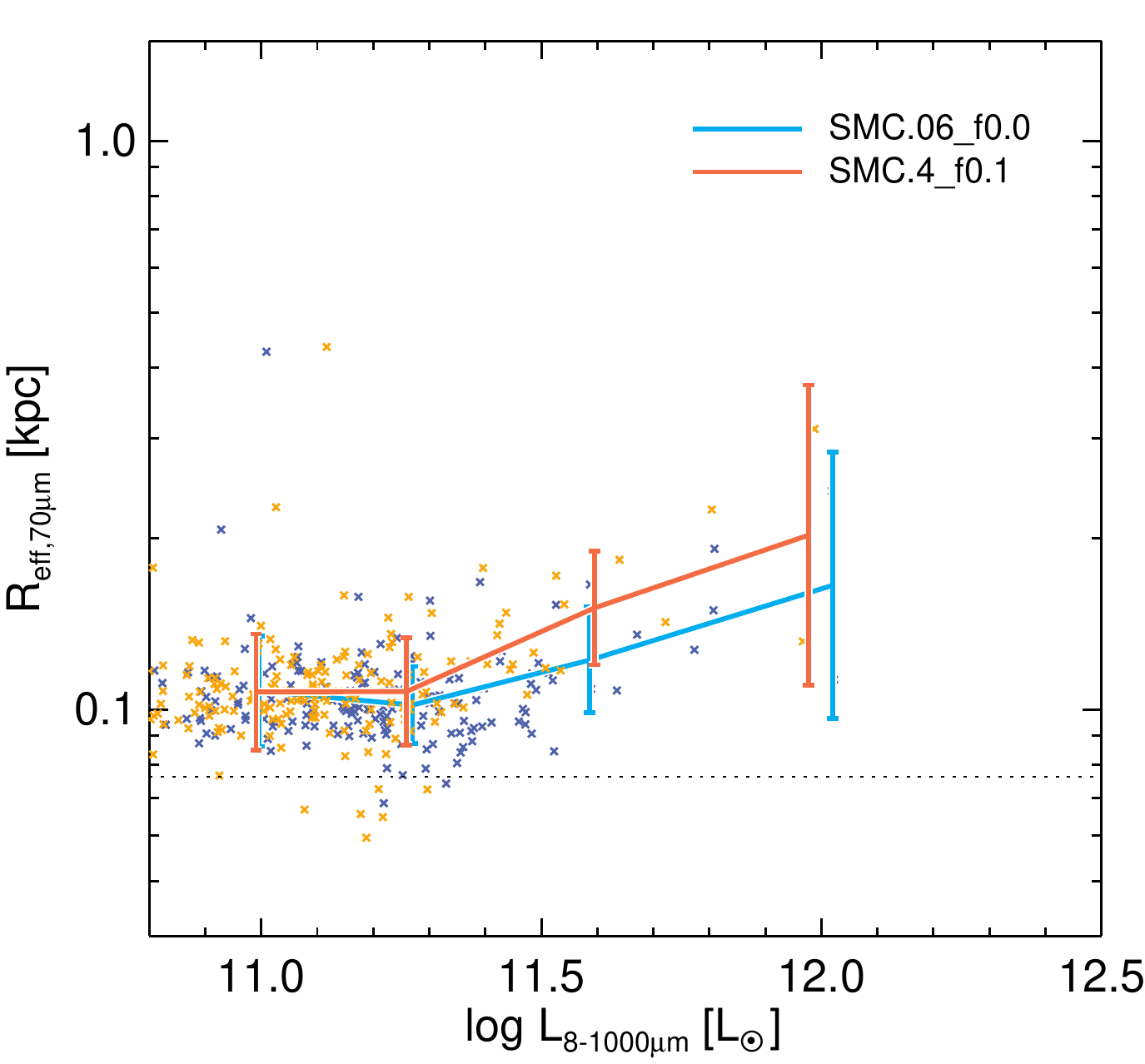}} 
\vskip -0cm
\caption{%\footnotesize %\scriptsize
Left:  FUV effective radius measured after smoothing with a Gaussian of FWHM=0.1 kpc
 as a function of GALEX FUV magnitude for simulated galaxies at $z=7$.
Included as black symbols are observational estimates by 
\citet[][circles]{2012Grazian}
and 
\citet[][triangles]{2012Ono}.
Right:  effective radius in FIR as a function of FIR luminosity for galaxies at $z=7$.
The effective radius is measured on restframe MIPS 70 $\mu$m band images after smoothing with a Gaussian of FWHM=0.1 kpc.
The dotted line marks the effective resolution of our simulation (75 pc). 
The solid lines and error bars indicate the mean and standard deviation.
}
\label{fig:re}
\end{figure}

We examine the sizes of simulated $z=7$ galaxies.
We first examine UV sizes.
The left panel of Figure~\ref{fig:re} shows the effective radius in GALEX FUV band as a function of FUV magnitude.
We see that the simulated galaxies have somewhat smaller sizes than the observed counterparts in the rest-frame FUV band,
although the trend that more UV luminous galaxies have larger sizes is in agreement with observations and
there is significant overalap between observations and our simulation.
The red curve with $f_{\rm esc}=10\%$ is 
actually computed not including the 10\% directly escaped light,
because we are unsure how to best model it without introducing some additional unchecked  
parameters.
We now turn to FIR sizes of simulated $z=7$ galaxies, 
shown in the right panel of Figure~\ref{fig:re}.
What is striking is that the predicted sizes of $z=7$ galaxies in the FIR band 
are very small, with effective radii being in the range of $80-500$pc proper.
The effective resolution of our simulation is $75$pc [$1.8$ cells, see \citet{2013bCen}],
which is probably the cause of the size floor seen.
Thus, we expect the lower mass galaxies may have effective radii in FIR 
that are smaller than $\sim 80$pc.
Nevertheless, we expect that the sizes in the range $100-500$pc predicted to be
real, to the extent that our simulation resolution is adeqaute for resolving them.
Because the FIR emission is primarily a function of the metal density distribution in the simulated galaxies,
they are much less prone to attenuation effects and hence are more robust.
However, there are possible caveats that the reader should keep in mind.
The central star formation may be dependent on feedback processes from star formation.
Since the exact strength of stellar feedback depends on an array of factors
that each have significant uncertainties, including the initial stellar mass function (IMF),
supernova feedback prescription used in the simulation, porosity of the interstellar medium
in the simulation that may be underestimated, feedback from AGN that is not included, etc,
our best estimate is that the sizes of the simulated galaxies may have been underestimated somewhat presently.
A comparison between our simulated results and upcoming observations will thus
shed useful light on both the cosmological model and astrophysics of galaxy formation at high redshift.

%\begin{figure}[!hb]
\begin{figure}[!h]    %Figure 4
\centering
\vskip -0cm
%\resizebox{5.0in}{!}{\includegraphics[angle=0]{ssfr_SIR.pdf}} 
\resizebox{3.5in}{!}{\includegraphics[angle=0]{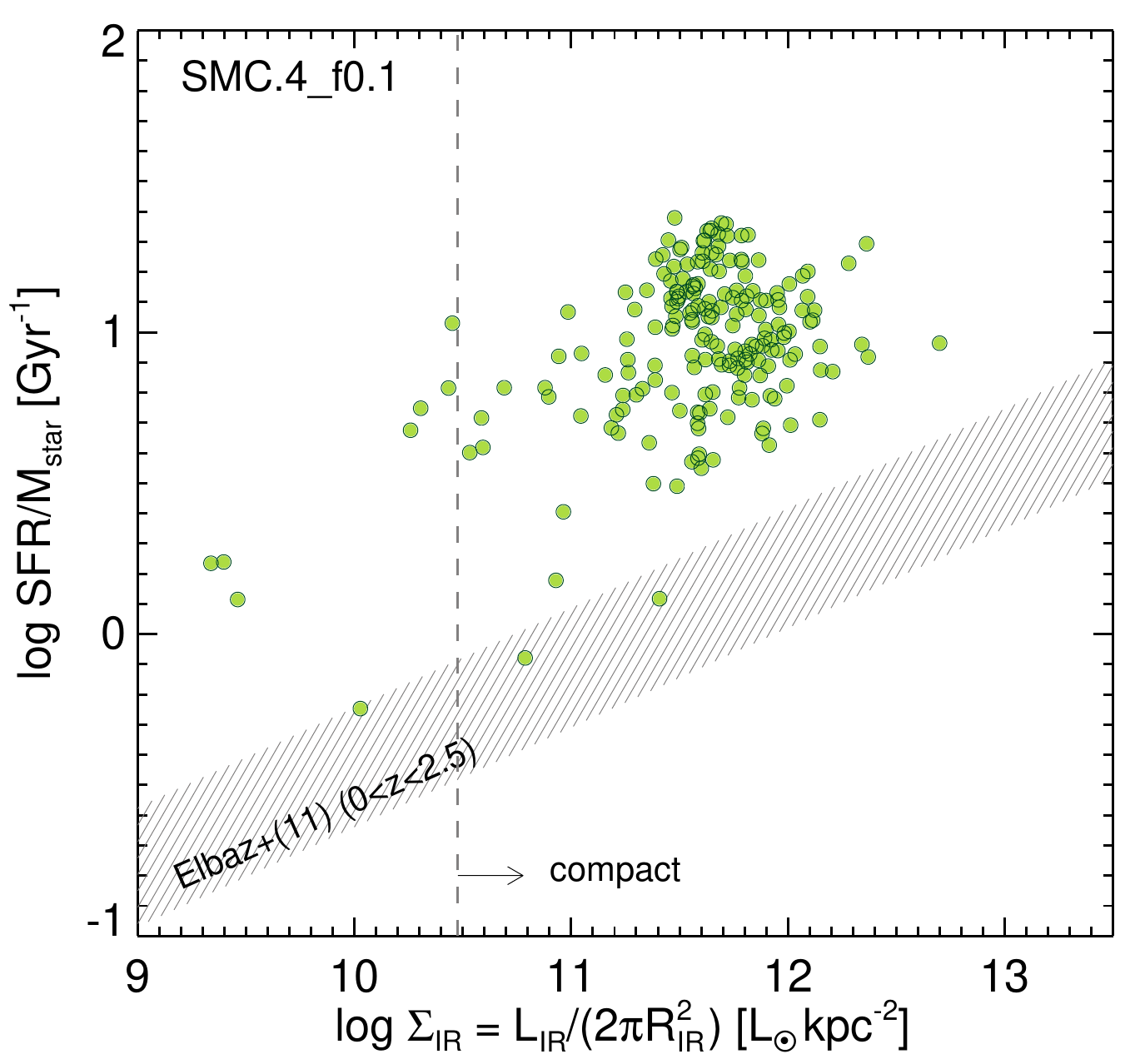}} 
\hskip -0cm
\caption{%\footnotesize %\scriptsize
shows the sSFR as a function of FIR surface brightness ($\Sigma_{\rm IR}$) 
for simulated galaxies at $z=7$ (circles).
$\Sigma_{\rm IR}$ is computed as $L_{8-1000\mu m} / 2\pi R_{IR}^2$, where $R_{IR}$ is the effective radius in the restframe MIPS 70 $\mu$m band.
Only one dust model is shown, because $\Sigma_{\rm IR}$ is insensitive to details of the dust model.
For comparisons, also shown as the shaded region is the star-forming sequence from the GOODS-Herschel sample 
at $0<z<2.5$ \citep[][]{2011Elbaz}.
}
\label{fig:ssfr_SIR}
\end{figure}

Another way to demonstrate the concentration of star formation in $z=7$ galaxies
is to plot the specific star formation rate as a function of FIR surface brightness,
shown in  Figure~\ref{fig:ssfr_SIR}. Comparing to the galaxies in the star-forming sequence 
at moderate redshift ($0<z<2.5$) \citep[][]{2011Elbaz},
it is evident that our simulated galaxies at $z=7$ 
are substantially more star-bursting by a factor of $3-20$ at a 
fixed surface density. 
\citet{2011Elbaz} also defines a galaxy as ``compact'' if
the FIR surface brightness ($\Sigma_{\rm IR}=L_{\rm IR}/2\pi R_{\rm IR}^2$) is 
greater than $3\times10^{10}\,L_{\odot}\,$kpc$^{-2}$. This roughly corresponds 
to galaxies emitting more than 60\% of their 13.2 $\mu m$ flux in the unresolved 
central $\sim 3\,{\rm kpc^2}$ region \citep[see][]{2010DiazSantos}.
According to the definition, the majority of the simulated galaxies will be classified as compact galaxies.

\begin{figure}[!ht]  %Figure 5
\centering
\vskip -0cm
\hskip -1cm
\resizebox{3.4in}{!}{\includegraphics[angle=0]{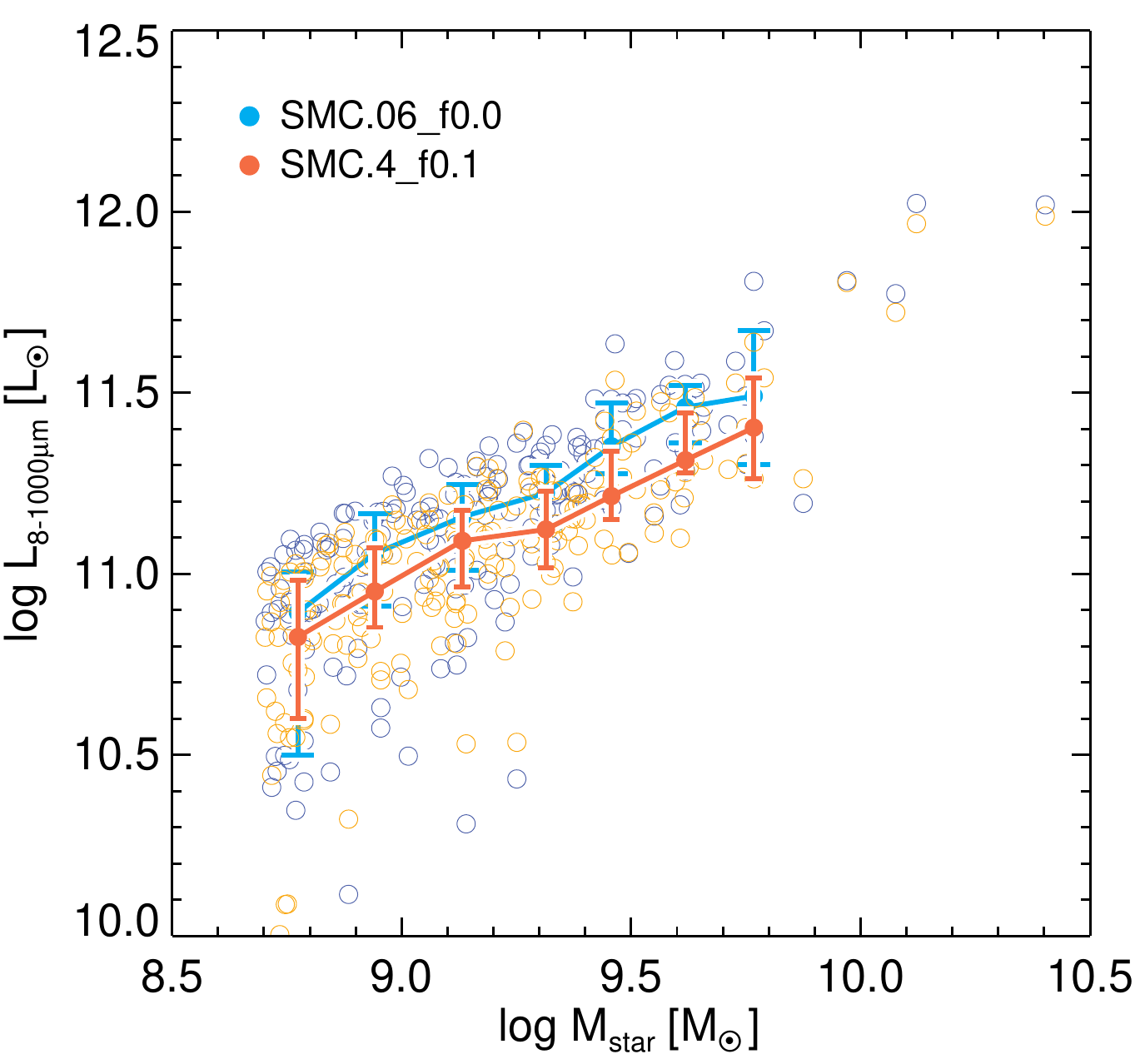}} 
\hskip -0cm
\resizebox{3.4in}{!}{\includegraphics[angle=0]{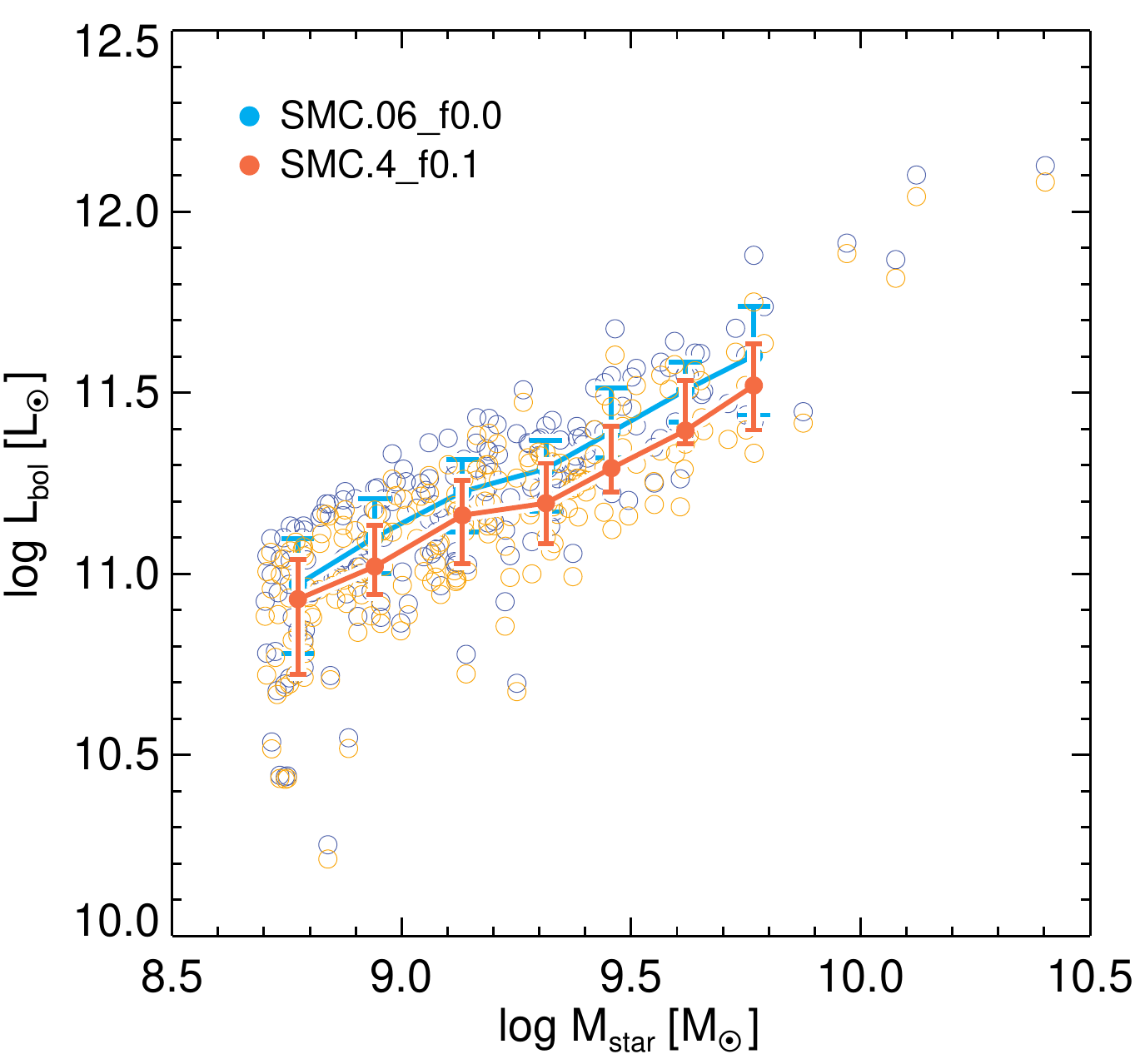}} 
\vskip -0cm
\caption{%\footnotesize %\scriptsize
shows FIR (left) and bolometric (right) luminosities in the range 8--1000$\mu$m of each galaxy at $z=7$ 
as a function of the galaxy stellar mass,
for two models with ($D/M$, $f_{\rm esc}$), (0.4,0.1) (red open circles) and (0.06,0) (blue open circles).
Solid points with error bars represent the median and interquartile range.
}
\label{fig:Mstar_IR}
\end{figure}

\begin{figure}[ht]    %Figure 6
\centering
\vskip -0.0cm
\hskip -1cm
\resizebox{3.4in}{!}{\includegraphics[angle=0]{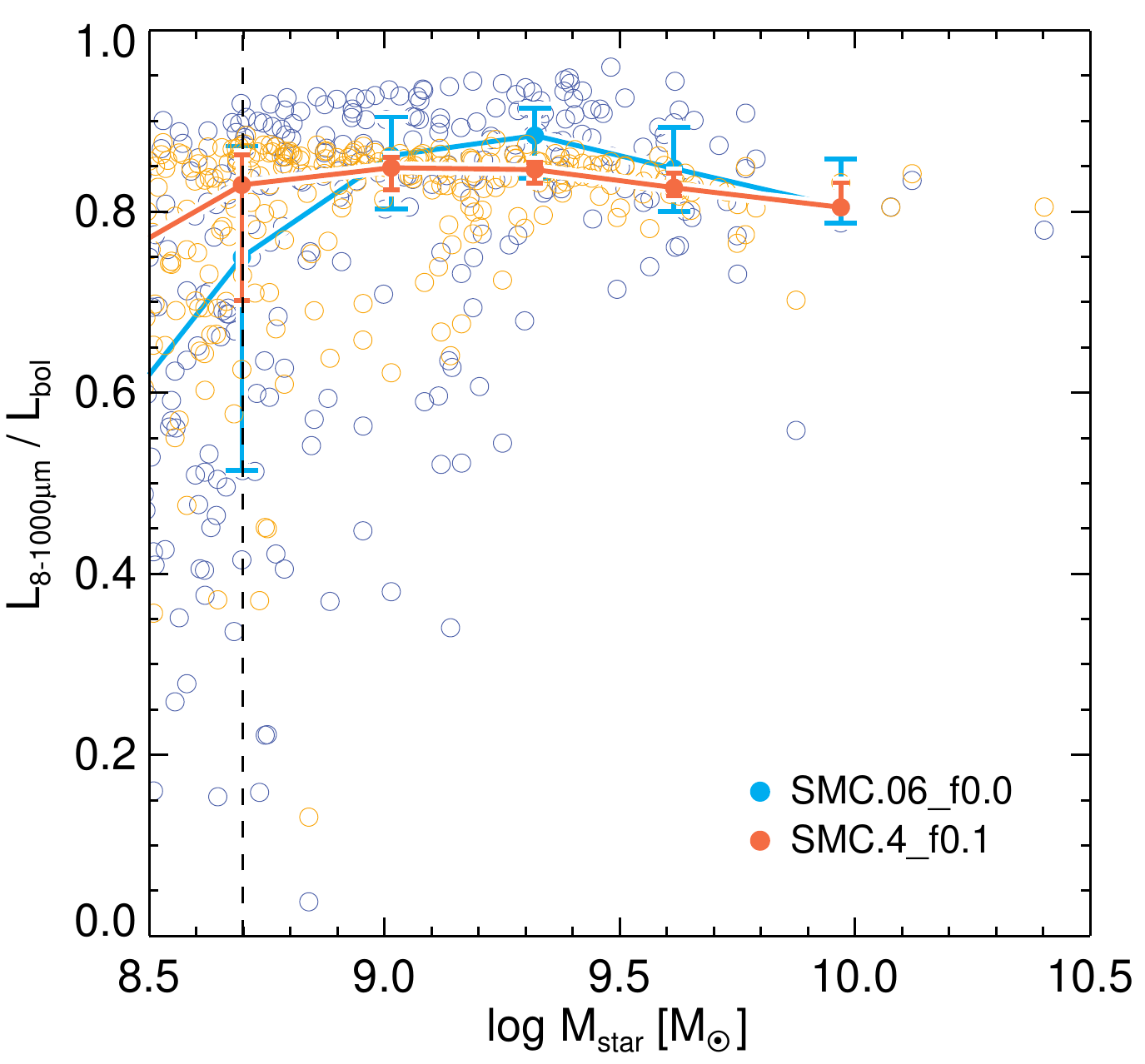}} 
\hskip -0cm
\resizebox{3.4in}{!}{\includegraphics[angle=0]{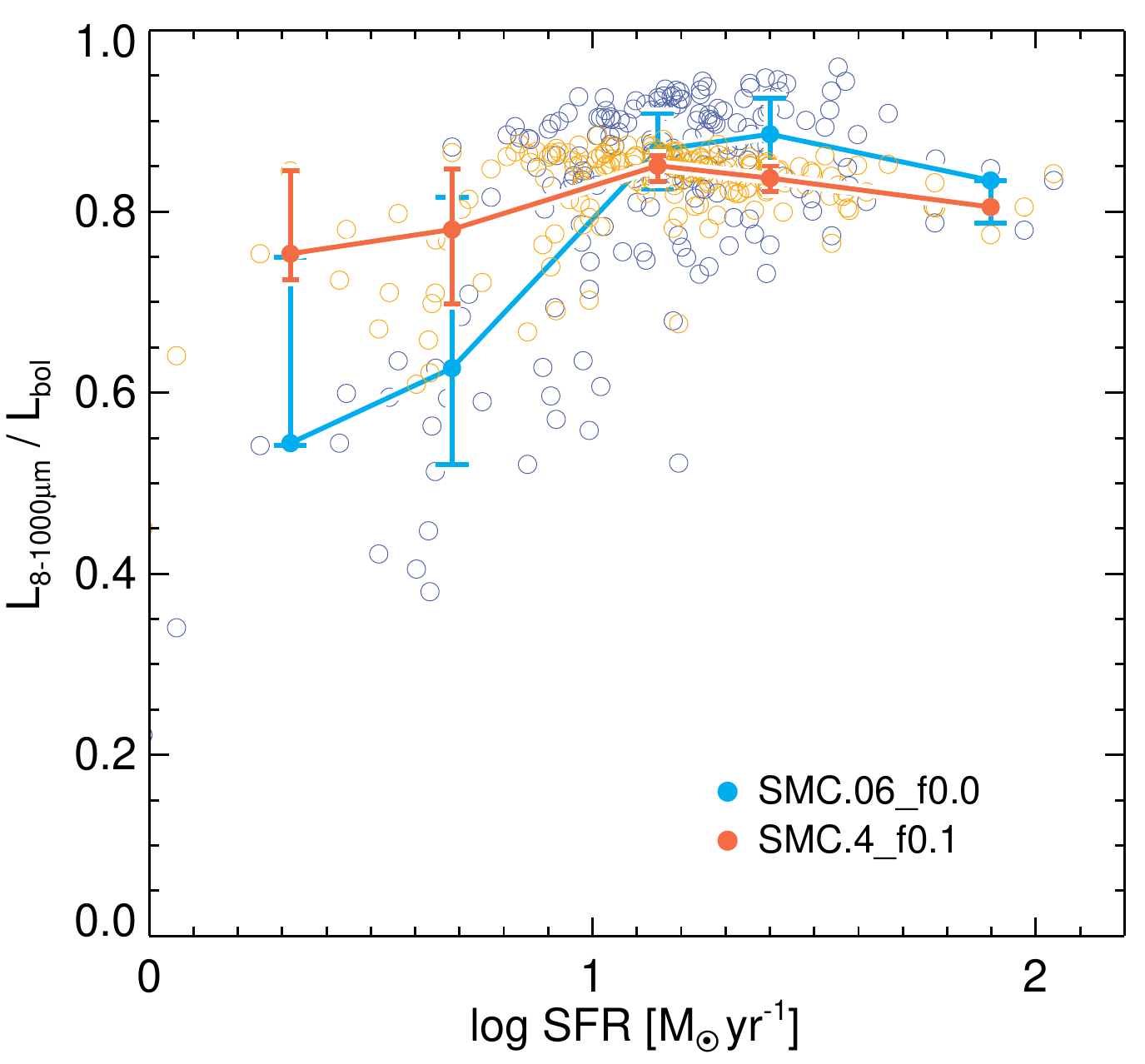}} 
\vskip -0cm
\caption{%\footnotesize %\scriptsize
The ratio of FIR luminosity appearing in the range $8-1000\mu m$ to the bolometric luminosity
as a function of stellar mass (left) and star formation rate (right). Blue and red circles denote the model
   with a dust-to-metal ratio ($D/M$) of 0.06 and 0.4, respectively, as indicated in the legend.
   An SMC-type dust extinctin curve is used in both cases. Points with error bars represent median
   and interquartile range.
}
\label{fig:rat}
\end{figure}

Having examined the SEDs and sizes, our attention is shifted to
the FIR radiation output of $z=7$ galaxies relative to their bolometric luminosities.
Figure~\ref{fig:Mstar_IR}
shows FIR in the range 8--1000$\mu$m (left panel) and bolometric (right panel) luminosities 
of each galaxy at $z=7$ 
as a function of the galaxy stellar mass,
for two models with ($D/M$, $f_{\rm esc}$), (0.4,0.1) (red open circles) and (0.06,0) (blue open circles).
Let us first study the right panel.
It is evident that bolometric luminosity (i.e., SFR) scales with stellar mass substantially sublinearly,
$L_{\rm bol}\propto M_{\rm star}^\xi$ with $\xi\sim 0.6-0.8$.
We attribute this, primarily, to ``stochasticity'' of star formation, where galaxies with higher specific star formation 
are in some ``bursting periods'', while those with relatively lower specific star formation rates presently 
are in relatively quiet ``patches''.
This explanation is consistent with the $\sim 0.5$dex spread in the bolometric luminosity at a fixed stellar mass.
The relation between FIR luminosity and stellar mass (left panel) is largely inherited from that of 
between bolometric luminosity and stellar mass (right panel),
with a small correction due to a non-uniform ratio of FIR to bolometric luminosities, shown in Figure~\ref{fig:rat} below.
If confirmed by ALMA, this will provide significant insight into the galaxy formation process at high redshift.
The expectation is that, with time, galaxies 
will mature and become stellar mass dominated on the baryon budget, and as such, the overall stochasticity 
is expected to decline towards lower redshift, resulting in $\xi$ gradually approaching unity.

Figure~\ref{fig:rat} shows the ratio of FIR luminosity in the range $8-1000\mu m$ to the bolometric luminosity 
for each galaxy as a function of stellar mass (left) and star formation rate (right), 
for two models  with a dust-to-metal ratio ($D/M$) of 0.06 and 0.4.
It is seen that there is a tendency for less massive galaxies or lower SFR galaxies to have lower median ratios.
For galaxies with SFR$=3-100\msun$/yr, it is found that the median FIR/bolometric luminosity ratio is in the range of $60-90\%$.
For galaxies with stellar mass in the range of $10^9-10^{10.5}\msun$, 
the median FIR/bolometric luminosity ratio is predicted to be in the range of $80-90\%$.
We expect that targeted observations of current high redshift galaxies at $z\sim 7$
by ALMA should be able to verify these predictions.

\begin{figure}[!ht]
\centering
\vskip 1cm
\resizebox{5.0in}{!}{\includegraphics[angle=0]{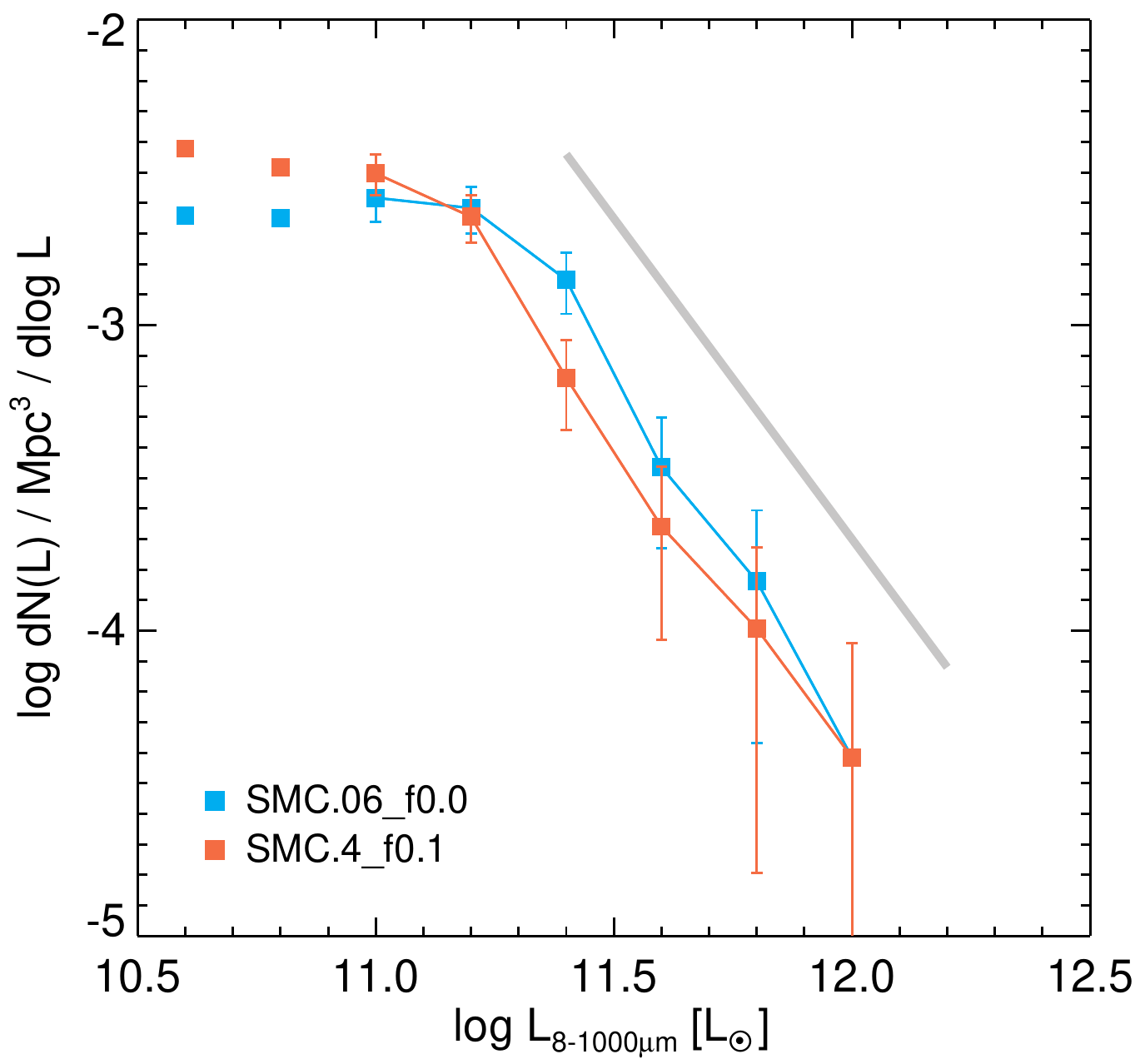}} 
\vskip -0cm
\caption{%\footnotesize %\scriptsize
The infrared luminosity function at $z=7$. Different color-codings correspond to models with
   a different assumption on $D/M$, as indicated in the legend. Solid lines display our reliable estimates for the number
   density, which can be directly compared to future observations. Other two bins below
   $ L_{8-1000\mu m} = 10^{11} L_\odot$ contain more than 20\% of under-resolved galaxies of stellar mass
   $M_{\rm star} < 2\times10^8 M_\odot$.  
$1\sigma$ error bars shown are Poissonian. 
The simulated LFs can be fit by a power law [$\log dN \propto (-2.1\pm0.4)d\log L$], shown as the gray line.
}
\label{fig:IRLF}
\end{figure}

Figure~\ref{fig:IRLF} shows the galaxy FIR luminosity function at $z=7$.
It is seen that the FIR luminosity function only depends weakly on the changes of dust properties;
the two dust models that are found to be able to match UV/optical properties of galaxies at $z=7$
are shown to give similar results.
Interestingly, for the high end luminosity range probed, $L_{\rm FIR}=10^{11}-10^{12}\lsun$,
there is no indication of an exponential drop that is normally the case for galaxy stellar mass function 
or UV/optical luminosity function at high redshift.
We find that a power-law slope of $-2.1\pm0.4$ provides a good fit to our simulated FIR luminosity functions,
where the uncertainty is measured by bootstrap resampling.
This power-law shape is related to the stochasticity of star formation of $z=7$ galaxies,
as noted above, which will have significant implications on small-scale power in the standard cold dark matter model
as well as the consumption and thermodynamic state of the gas prior to star formation in galaxies at $z=7$.
It is expected that observing the UV selected high redshift galaxy candidates in the UDF
by ALMA may be able to check this expected power-law slope.
It is noted that, since the ratio of FIR to bolometric luminosity is close to unity (see Figure~\ref{fig:rat}),
this power-law behavior is rather robust to 
attentuation uncertainties that plague rest-frame UV luminosity functions,
making FIR continuum observations by ALMA especially powerful.
If verified, we will learn a great deal of the mode of star formation in high-z galaxies:
how bursty is star formation in high redshift galaxies
and what is the dispersion in star formation rate at a fixed halo or stellar mass?

\section{Conclusions}

We perform, in unprecedented details, three-dimensional panchromatic dust radiative transfer calculations
on a set of 198 galaxies of stellar masses in the range $5\times 10^8-3\times10^{10}\msun$ resolved at $29$h$^{-1}$pc at $z\sim 7$ from 
an {\it ab initio} cosmological hydrodynamic simulation of the standard cold dark matter model.
The soundness of treatment of relevant physical processes has been checked by comparing to a set of independent observations.
In a companion paper \citep{2013Kimm} we show that 
the UV-optical properties, including stellar mass and luminosity functions, UV-optical and FUV-NUV colors,
are in good agreement with observations, if an SMC-type dust extinction curve is adopted.
In order to make further testable predictions of the same model,
we present here, self-consistently, an additional set of infrared continuum properties of these $z\sim 7$ simulated galaxies,
to be confronted with upcoming ALMA data.
The main results may be summarized in four points.

(1) The effective radius in the restframe MIPS$70\mu$m band is in the range of $80-400$pc proper for $z=7$ galaxies with $L_{\rm FIR}=10^{11.3-12}\lsun$,
corresponding to an angular size of $0^{''}.015-0^{''}.075$, which may be resolvable by ALMA.
The majority of the simulated galaxies would be classified as compact starburst galaxies \citep{2011Elbaz}.
The predicted size would be below the extrapolation from lower redshift evolution in rest-frame UV or visual band \citep[e.g.,][]{2004Ferguson, 2006Trujillo}.
%(1+z)-0.45+/-0.10.

(2) The median of the peak wavelength of the FIR SED is in the range of $45-60\mu$m, depending on the dust-to-metal ratio,
corresponding to observed wavelength of $360-480\mu$m, within the ALMA range.
The peak wavelength found is comparable to local starburst galaxies, such as Arp 220 and NGC 6240.

(3) For galaxies with SFR in the range $3-100\msun$~yr$^{-1}$, the median FIR in the range 8--1000$\mu$m to bolometric luminosity ratio is $60-90\%$.
In other words, the SFR inferred
based on the observed optical and UV luminosities is missing more than one half of the total SFR, on average. 
This is a new and key prediction that should be easily verifiable by upcoming ALMA observations, in conjunction with HST UDF data.

(4) The FIR luminosity function displays a power law in the high end with a slope of -3.1$\pm$0.4, instead of the usual exponential decline.
A related result is that $L_{\rm bol}\propto M_{\rm star}^\xi$ with $\xi\sim 0.6-0.8$,
which is somewhat flatter than the corresponding relation at lower redshifts.
Both results may be attributable to ``stochasticity" and significant dispersions in star formation at a given halo mass.

\section*{Acknowledgements}
We would like to thank Bruce Draine, Simona Gallerani, Roberto Maiolino and John Wise for discussion,
and an anonymous referee for a very useful constructive report.
Computing resources were in part provided by the NASA High-
End Computing (HEC) Program through the NASA Advanced
Supercomputing (NAS) Division at Ames Research Center.
The research is supported in part by NSF grant AST-1108700 and NASA grant NNX12AF91G.
Some of our analysis is done using the {\sc YT} toolkit \citep{turk11}.

%\bibliographystyle{apj}
%\bibliography{astro}

\begin{thebibliography}{53}
\expandafter\ifx\csname natexlab\endcsname\relax\def\natexlab#1{#1}\fi

\bibitem[{{Abel} {et~al.}(1997){Abel}, {Anninos}, {Zhang}, \&
  {Norman}}]{1997Abel}
{Abel}, T., {Anninos}, P., {Zhang}, Y., \& {Norman}, M.~L. 1997, New Astronomy,
  2, 181

\bibitem[{{Bouwens} {et~al.}(2012){Bouwens}, {Illingworth}, {Oesch}, {Franx},
  {Labb{\'e}}, {Trenti}, {van Dokkum}, {Carollo}, {Gonz{\'a}lez}, {Smit}, \&
  {Magee}}]{2012aBouwens}
{Bouwens}, R.~J., {Illingworth}, G.~D., {Oesch}, P.~A., {Franx}, M.,
  {Labb{\'e}}, I., {Trenti}, M., {van Dokkum}, P., {Carollo}, C.~M.,
  {Gonz{\'a}lez}, V., {Smit}, R., \& {Magee}, D. 2012, \apj, 754, 83

\bibitem[{{Bouwens} {et~al.}(2010){Bouwens}, {Illingworth}, {Oesch}, {Labbe},
  {Trenti}, {van Dokkum}, {Franx}, {Stiavelli}, {Carollo}, {Magee}, \&
  {Gonzalez}}]{2010eBouwens}
{Bouwens}, R.~J., {Illingworth}, G.~D., {Oesch}, P.~A., {Labbe}, I., {Trenti},
  M., {van Dokkum}, P., {Franx}, M., {Stiavelli}, M., {Carollo}, C.~M.,
  {Magee}, D., \& {Gonzalez}, V. 2010, ArXiv e-prints

\bibitem[{{Bruzual} \& {Charlot}(2003)}]{Bruzual03}
{Bruzual}, G., \& {Charlot}, S. 2003, \mnras, 344, 1000

\bibitem[{{Bryan}(1999)}]{1999aBryan}
{Bryan}, G.~L. 1999, Comput.~Sci.~Eng., Vol.~1, No.~2, p.~46 - 53, 1, 46

\bibitem[{{Bunker} {et~al.}(2010){Bunker}, {Wilkins}, {Ellis}, {Stark},
  {Lorenzoni}, {Chiu}, {Lacy}, {Jarvis}, \& {Hickey}}]{2010Bunker}
{Bunker}, A.~J., {Wilkins}, S., {Ellis}, R.~S., {Stark}, D.~P., {Lorenzoni},
  S., {Chiu}, K., {Lacy}, M., {Jarvis}, M.~J., \& {Hickey}, S. 2010, \mnras,
  1378

\bibitem[{{Carilli} {et~al.}(2008){Carilli}, {Walter}, {Wang}, {Wootten},
  {Menten}, {Bertoldi}, {Schinnerer}, {Cox}, {Beelen}, \&
  {Omont}}]{2008Carilli}
{Carilli}, C.~L., {Walter}, F., {Wang}, R., {Wootten}, A., {Menten}, K.,
  {Bertoldi}, F., {Schinnerer}, E., {Cox}, P., {Beelen}, A., \& {Omont}, A.
  2008, \apss, 313, 307

\bibitem[{{Cen}(2003)}]{2003Cen}
{Cen}, R. 2003, \apj, 591, 12

\bibitem[{{Cen}(2012)}]{2012Cen}
---. 2012, \apj, 748, 121

\bibitem[{{Cen}(2013)}]{2013bCen}
---. 2013, ArXiv e-prints

\bibitem[{{Cen} {et~al.}(1995){Cen}, {Kang}, {Ostriker}, \& {Ryu}}]{1995Cen}
{Cen}, R., {Kang}, H., {Ostriker}, J.~P., \& {Ryu}, D. 1995, \apj, 451, 436

\bibitem[{{Cen} {et~al.}(2005){Cen}, {Nagamine}, \& {Ostriker}}]{2005Cen}
{Cen}, R., {Nagamine}, K., \& {Ostriker}, J.~P. 2005, \apj, 635, 86

\bibitem[{{Cen} \& {Ostriker}(1992)}]{1992CenOstriker}
{Cen}, R., \& {Ostriker}, J.~P. 1992, \apjl, 399, L113

\bibitem[{{Dalgarno} \& {McCray}(1972)}]{1972Dalgarno}
{Dalgarno}, A., \& {McCray}, R.~A. 1972, \araa, 10, 375

\bibitem[{{D{\'{\i}}az-Santos} {et~al.}(2010){D{\'{\i}}az-Santos},
  {Charmandaris}, {Armus}, {Petric}, {Howell}, {Murphy}, {Mazzarella},
  {Veilleux}, {Bothun}, {Inami}, {Appleton}, {Evans}, {Haan}, {Marshall},
  {Sanders}, {Stierwalt}, \& {Surace}}]{2010DiazSantos}
{D{\'{\i}}az-Santos}, T., {Charmandaris}, V., {Armus}, L., {Petric}, A.~O.,
  {Howell}, J.~H., {Murphy}, E.~J., {Mazzarella}, J.~M., {Veilleux}, S.,
  {Bothun}, G., {Inami}, H., {Appleton}, P.~N., {Evans}, A.~S., {Haan}, S.,
  {Marshall}, J.~A., {Sanders}, D.~B., {Stierwalt}, S., \& {Surace}, J.~A.
  2010, \apj, 723, 993

\bibitem[{{Dopita} {et~al.}(2005){Dopita}, {Groves}, {Fischera}, {Sutherland},
  {Tuffs}, {Popescu}, {Kewley}, {Reuland}, \& {Leitherer}}]{dopita05}
{Dopita}, M.~A., {Groves}, B.~A., {Fischera}, J., {Sutherland}, R.~S., {Tuffs},
  R.~J., {Popescu}, C.~C., {Kewley}, L.~J., {Reuland}, M., \& {Leitherer}, C.
  2005, \apj, 619, 755

\bibitem[{{Draine} \& {Li}(2007)}]{draine07}
{Draine}, B.~T., \& {Li}, A. 2007, \apj, 657, 810

\bibitem[{{Dunlop} {et~al.}(2012){Dunlop}, {McLure}, {Robertson}, {Ellis},
  {Stark}, {Cirasuolo}, \& {de Ravel}}]{2012Dunlop}
{Dunlop}, J.~S., {McLure}, R.~J., {Robertson}, B.~E., {Ellis}, R.~S., {Stark},
  D.~P., {Cirasuolo}, M., \& {de Ravel}, L. 2012, \mnras, 420, 901

\bibitem[{{Dunlop} {et~al.}(2013){Dunlop}, {Rogers}, {McLure}, {Ellis},
  {Robertson}, {Koekemoer}, {Dayal}, {Curtis-Lake}, {Wild}, {Charlot},
  {Bowler}, {Schenker}, {Ouchi}, {Ono}, {Cirasuolo}, {Furlanetto}, {Stark},
  {Targett}, \& {Schneider}}]{2013Dunlop}
{Dunlop}, J.~S., {Rogers}, A.~B., {McLure}, R.~J., {Ellis}, R.~S., {Robertson},
  B.~E., {Koekemoer}, A., {Dayal}, P., {Curtis-Lake}, E., {Wild}, V.,
  {Charlot}, S., {Bowler}, R.~A.~A., {Schenker}, M.~A., {Ouchi}, M., {Ono}, Y.,
  {Cirasuolo}, M., {Furlanetto}, S.~R., {Stark}, D.~P., {Targett}, T.~A., \&
  {Schneider}, E. 2013, \mnras, 432, 3520

\bibitem[{Eisenstein \& Hu(1999)}]{1999Eisenstein}
Eisenstein, D., \& Hu, P. 1999, ApJ, 511, 5

\bibitem[{{Elbaz} {et~al.}(2011){Elbaz}, {Dickinson}, {Hwang},
  {D{\'{\i}}az-Santos}, {Magdis}, {Magnelli}, {Le Borgne}, {Galliano},
  {Pannella}, {Chanial}, {Armus}, {Charmandaris}, {Daddi}, {Aussel}, {Popesso},
  {Kartaltepe}, {Altieri}, {Valtchanov}, {Coia}, {Dannerbauer}, {Dasyra},
  {Leiton}, {Mazzarella}, {Alexander}, {Buat}, {Burgarella}, {Chary}, {Gilli},
  {Ivison}, {Juneau}, {Le Floc'h}, {Lutz}, {Morrison}, {Mullaney}, {Murphy},
  {Pope}, {Scott}, {Brodwin}, {Calzetti}, {Cesarsky}, {Charlot}, {Dole},
  {Eisenhardt}, {Ferguson}, {F{\"o}rster Schreiber}, {Frayer}, {Giavalisco},
  {Huynh}, {Koekemoer}, {Papovich}, {Reddy}, {Surace}, {Teplitz}, {Yun}, \&
  {Wilson}}]{2011Elbaz}
{Elbaz}, D., {Dickinson}, M., {Hwang}, H.~S., {D{\'{\i}}az-Santos}, T.,
  {Magdis}, G., {Magnelli}, B., {Le Borgne}, D., {Galliano}, F., {Pannella},
  M., {Chanial}, P., {Armus}, L., {Charmandaris}, V., {Daddi}, E., {Aussel},
  H., {Popesso}, P., {Kartaltepe}, J., {Altieri}, B., {Valtchanov}, I., {Coia},
  D., {Dannerbauer}, H., {Dasyra}, K., {Leiton}, R., {Mazzarella}, J.,
  {Alexander}, D.~M., {Buat}, V., {Burgarella}, D., {Chary}, R.-R., {Gilli},
  R., {Ivison}, R.~J., {Juneau}, S., {Le Floc'h}, E., {Lutz}, D., {Morrison},
  G.~E., {Mullaney}, J.~R., {Murphy}, E., {Pope}, A., {Scott}, D., {Brodwin},
  M., {Calzetti}, D., {Cesarsky}, C., {Charlot}, S., {Dole}, H., {Eisenhardt},
  P., {Ferguson}, H.~C., {F{\"o}rster Schreiber}, N., {Frayer}, D.,
  {Giavalisco}, M., {Huynh}, M., {Koekemoer}, A.~M., {Papovich}, C., {Reddy},
  N., {Surace}, C., {Teplitz}, H., {Yun}, M.~S., \& {Wilson}, G. 2011, \aap,
  533, A119

\bibitem[{{Fan} {et~al.}(2006){Fan}, {Strauss}, {Becker}, {White}, {Gunn},
  {Knapp}, {Richards}, {Schneider}, {Brinkmann}, \& {Fukugita}}]{2006Fan}
{Fan}, X., {Strauss}, M.~A., {Becker}, R.~H., {White}, R.~L., {Gunn}, J.~E.,
  {Knapp}, G.~R., {Richards}, G.~T., {Schneider}, D.~P., {Brinkmann}, J., \&
  {Fukugita}, M. 2006, \aj, 132, 117

\bibitem[{{Faucher-Gigu{\`e}re} {et~al.}(2008){Faucher-Gigu{\`e}re}, {Lidz},
  {Hernquist}, \& {Zaldarriaga}}]{2008FG}
{Faucher-Gigu{\`e}re}, C., {Lidz}, A., {Hernquist}, L., \& {Zaldarriaga}, M.
  2008, \apj, 688, 85

\bibitem[{Ferguson {et~al.}(2004)Ferguson, Dickinson, Giavalisco, Kretchmer,
  Ravindranath, Idzi, Taylor, Conselice, {et~al.}}]{2004Ferguson}
Ferguson, H.~C., Dickinson, M., Giavalisco, M., Kretchmer, C., Ravindranath,
  S., Idzi, R., Taylor, E., Conselice, C.~J., {et~al.} 2004, ApJ, 600, L107

\bibitem[{{Finkelstein} {et~al.}(2012){Finkelstein}, {Papovich}, {Salmon},
  {Finlator}, {Dickinson}, {Ferguson}, {Giavalisco}, {Koekemoer}, {Reddy},
  {Bassett}, {Conselice}, {Dunlop}, {Faber}, {Grogin}, {Hathi}, {Kocevski},
  {Lai}, {Lee}, {McLure}, {Mobasher}, \& {Newman}}]{2012Finkelstein}
{Finkelstein}, S.~L., {Papovich}, C., {Salmon}, B., {Finlator}, K.,
  {Dickinson}, M., {Ferguson}, H.~C., {Giavalisco}, M., {Koekemoer}, A.~M.,
  {Reddy}, N.~A., {Bassett}, R., {Conselice}, C.~J., {Dunlop}, J.~S., {Faber},
  S.~M., {Grogin}, N.~A., {Hathi}, N.~P., {Kocevski}, D.~D., {Lai}, K., {Lee},
  K.-S., {McLure}, R.~J., {Mobasher}, B., \& {Newman}, J.~A. 2012, \apj, 756,
  164

\bibitem[{{Gonz{\'a}lez} {et~al.}(2011){Gonz{\'a}lez}, {Labb{\'e}}, {Bouwens},
  {Illingworth}, {Franx}, \& {Kriek}}]{2011Gonzalez}
{Gonz{\'a}lez}, V., {Labb{\'e}}, I., {Bouwens}, R.~J., {Illingworth}, G.,
  {Franx}, M., \& {Kriek}, M. 2011, \apjl, 735, L34

\bibitem[{{Grazian} {et~al.}(2012){Grazian}, {Castellano}, {Fontana},
  {Pentericci}, {Dunlop}, {McLure}, {Koekemoer}, {Dickinson}, {Faber},
  {Ferguson}, {Galametz}, {Giavalisco}, {Grogin}, {Hathi}, {Kocevski}, {Lai},
  {Newman}, \& {Vanzella}}]{2012Grazian}
{Grazian}, A., {Castellano}, M., {Fontana}, A., {Pentericci}, L., {Dunlop},
  J.~S., {McLure}, R.~J., {Koekemoer}, A.~M., {Dickinson}, M.~E., {Faber},
  S.~M., {Ferguson}, H.~C., {Galametz}, A., {Giavalisco}, M., {Grogin}, N.~A.,
  {Hathi}, N.~P., {Kocevski}, D.~D., {Lai}, K., {Newman}, J.~A., \& {Vanzella},
  E. 2012, \aap, 547, A51

\bibitem[{{Groves} {et~al.}(2008){Groves}, {Dopita}, {Sutherland}, {Kewley},
  {Fischera}, {Leitherer}, {Brandl}, \& {van Breugel}}]{groves08}
{Groves}, B., {Dopita}, M.~A., {Sutherland}, R.~S., {Kewley}, L.~J.,
  {Fischera}, J., {Leitherer}, C., {Brandl}, B., \& {van Breugel}, W. 2008,
  \apjs, 176, 438

\bibitem[{{Haardt} \& {Madau}(1996)}]{1996Haardt}
{Haardt}, F., \& {Madau}, P. 1996, \apj, 461, 20

\bibitem[{{Heckman}(2001)}]{2001Heckman}
{Heckman}, T.~M. 2001, in Astronomical Society of the Pacific Conference
  Series, Vol. 240, Gas and Galaxy Evolution, ed. J.~E. {Hibbard}, M.~{Rupen},
  \& J.~H. {van Gorkom}, 345

\bibitem[{{Hinshaw} {et~al.}(2012){Hinshaw}, {Larson}, {Komatsu}, {Spergel},
  {Bennett}, {Dunkley}, {Nolta}, {Halpern}, {Hill}, {Odegard}, {Page}, {Smith},
  {Weiland}, {Gold}, {Jarosik}, {Kogut}, {Limon}, {Meyer}, {Tucker}, {Wollack},
  \& {Wright}}]{2012Hinshaw}
{Hinshaw}, G., {Larson}, D., {Komatsu}, E., {Spergel}, D.~N., {Bennett}, C.~L.,
  {Dunkley}, J., {Nolta}, M.~R., {Halpern}, M., {Hill}, R.~S., {Odegard}, N.,
  {Page}, L., {Smith}, K.~M., {Weiland}, J.~L., {Gold}, B., {Jarosik}, N.,
  {Kogut}, A., {Limon}, M., {Meyer}, S.~S., {Tucker}, G.~S., {Wollack}, E., \&
  {Wright}, E.~L. 2012, ArXiv e-prints

\bibitem[{{Hodge} {et~al.}(2013){Hodge}, {Karim}, {Smail}, {Swinbank},
  {Walter}, {Biggs}, {Ivison}, {Weiss}, {Alexander}, {Bertoldi}, {Brandt},
  {Chapman}, {Coppin}, {Cox}, {Danielson}, {Dannerbauer}, {De Breuck},
  {Decarli}, {Edge}, {Greve}, {Knudsen}, {Menten}, {Rix}, {Schinnerer},
  {Simpson}, {Wardlow}, \& {van der Werf}}]{2013Hodge}
{Hodge}, J.~A., {Karim}, A., {Smail}, I., {Swinbank}, A.~M., {Walter}, F.,
  {Biggs}, A.~D., {Ivison}, R.~J., {Weiss}, A., {Alexander}, D.~M., {Bertoldi},
  F., {Brandt}, W.~N., {Chapman}, S.~C., {Coppin}, K.~E.~K., {Cox}, P.,
  {Danielson}, A.~L.~R., {Dannerbauer}, H., {De Breuck}, C., {Decarli}, R.,
  {Edge}, A.~C., {Greve}, T.~R., {Knudsen}, K.~K., {Menten}, K.~M., {Rix},
  H.-W., {Schinnerer}, E., {Simpson}, J.~M., {Wardlow}, J.~L., \& {van der
  Werf}, P. 2013, \apj, 768, 91

\bibitem[{{Jonsson}(2006)}]{jonsson06}
{Jonsson}, P. 2006, \mnras, 372, 2

\bibitem[{{Jonsson} {et~al.}(2010){Jonsson}, {Groves}, \& {Cox}}]{jonsson10}
{Jonsson}, P., {Groves}, B.~A., \& {Cox}, T.~J. 2010, \mnras, 403, 17

\bibitem[{{Joung} {et~al.}(2009){Joung}, {Cen}, \& {Bryan}}]{2009Joung}
{Joung}, M.~R., {Cen}, R., \& {Bryan}, G.~L. 2009, \apjl, 692, L1

\bibitem[{{Kimm} \& {Cen}(2013)}]{2013Kimm}
{Kimm}, T., \& {Cen}, R. 2013, ArXiv e-prints

\bibitem[{{Komatsu} {et~al.}(2010){Komatsu}, {Smith}, {Dunkley}, {Bennett},
  {Gold}, {Hinshaw}, {Jarosik}, {Larson}, {Nolta}, {Page}, {Spergel},
  {Halpern}, {Hill}, {Kogut}, {Limon}, {Meyer}, {Odegard}, {Tucker}, {Weiland},
  {Wollack}, \& {Wright}}]{2010Komatsu}
{Komatsu}, E., {Smith}, K.~M., {Dunkley}, J., {Bennett}, C.~L., {Gold}, B.,
  {Hinshaw}, G., {Jarosik}, N., {Larson}, D., {Nolta}, M.~R., {Page}, L.,
  {Spergel}, D.~N., {Halpern}, M., {Hill}, R.~S., {Kogut}, A., {Limon}, M.,
  {Meyer}, S.~S., {Odegard}, N., {Tucker}, G.~S., {Weiland}, J.~L., {Wollack},
  E., \& {Wright}, E.~L. 2010, ArXiv e-prints

\bibitem[{{Krauss} \& {Turner}(1995)}]{1995Krauss}
{Krauss}, L.~M., \& {Turner}, M.~S. 1995, General Relativity and Gravitation,
  27, 1137

\bibitem[{{Labb{\'e}} {et~al.}(2010){Labb{\'e}}, {Gonz{\'a}lez}, {Bouwens},
  {Illingworth}, {Franx}, {Trenti}, {Oesch}, {van Dokkum}, {Stiavelli},
  {Carollo}, {Kriek}, \& {Magee}}]{2010Labbe}
{Labb{\'e}}, I., {Gonz{\'a}lez}, V., {Bouwens}, R.~J., {Illingworth}, G.~D.,
  {Franx}, M., {Trenti}, M., {Oesch}, P.~A., {van Dokkum}, P.~G., {Stiavelli},
  M., {Carollo}, C.~M., {Kriek}, M., \& {Magee}, D. 2010, \apjl, 716, L103

\bibitem[{{Leitherer} {et~al.}(1999){Leitherer}, {Schaerer}, {Goldader},
  {Delgado}, {Robert}, {Kune}, {de Mello}, {Devost}, \&
  {Heckman}}]{leitherer99}
{Leitherer}, C., {Schaerer}, D., {Goldader}, J.~D., {Delgado}, R.~M.~G.,
  {Robert}, C., {Kune}, D.~F., {de Mello}, D.~F., {Devost}, D., \& {Heckman},
  T.~M. 1999, \apjs, 123, 3

\bibitem[{{McLure} {et~al.}(2011){McLure}, {Dunlop}, {de Ravel}, {Cirasuolo},
  {Ellis}, {Schenker}, {Robertson}, {Koekemoer}, {Stark}, \&
  {Bowler}}]{2011Mclure}
{McLure}, R.~J., {Dunlop}, J.~S., {de Ravel}, L., {Cirasuolo}, M., {Ellis},
  R.~S., {Schenker}, M., {Robertson}, B.~E., {Koekemoer}, A.~M., {Stark},
  D.~P., \& {Bowler}, R.~A.~A. 2011, \mnras, 418, 2074

\bibitem[{{Ono} {et~al.}(2012){Ono}, {Ouchi}, {Curtis-Lake}, {Schenker},
  {Ellis}, {McLure}, {Dunlop}, {Robertson}, {Koekemoer}, {Bowler}, {Rogers},
  {Schneider}, {Charlot}, {Stark}, {Shimasaku}, {Furlanetto}, \&
  {Cirasuolo}}]{2012Ono}
{Ono}, Y., {Ouchi}, M., {Curtis-Lake}, E., {Schenker}, M.~A., {Ellis}, R.~S.,
  {McLure}, R.~J., {Dunlop}, J.~S., {Robertson}, B.~E., {Koekemoer}, A.~M.,
  {Bowler}, R.~A.~A., {Rogers}, A.~B., {Schneider}, E., {Charlot}, S., {Stark},
  D.~P., {Shimasaku}, K., {Furlanetto}, S.~R., \& {Cirasuolo}, M. 2012, ArXiv
  e-prints

\bibitem[{{O'Shea} {et~al.}(2005){O'Shea}, {Abel}, {Whalen}, \&
  {Norman}}]{2005OShea}
{O'Shea}, B.~W., {Abel}, T., {Whalen}, D., \& {Norman}, M.~L. 2005, \apjl, 628,
  L5

\bibitem[{{Planck Collaboration} {et~al.}(2013){Planck Collaboration}, {Ade},
  {Aghanim}, {Armitage-Caplan}, {Arnaud}, {Ashdown}, {Atrio-Barandela},
  {Aumont}, {Baccigalupi}, {Banday}, \& et~al.}]{2013Planck}
{Planck Collaboration}, {Ade}, P.~A.~R., {Aghanim}, N., {Armitage-Caplan}, C.,
  {Arnaud}, M., {Ashdown}, M., {Atrio-Barandela}, F., {Aumont}, J.,
  {Baccigalupi}, C., {Banday}, A.~J., \& et~al. 2013, ArXiv e-prints

\bibitem[{{Razoumov} \& {Sommer-Larsen}(2010)}]{2010Razoumov}
{Razoumov}, A.~O., \& {Sommer-Larsen}, J. 2010, \apj, 710, 1239

\bibitem[{{Stark} {et~al.}(2010){Stark}, {Ellis}, {Chiu}, {Ouchi}, \&
  {Bunker}}]{2010Stark}
{Stark}, D.~P., {Ellis}, R.~S., {Chiu}, K., {Ouchi}, M., \& {Bunker}, A. 2010,
  \mnras, 408, 1628

\bibitem[{{Trujillo} {et~al.}(2006){Trujillo}, {F{\"o}rster Schreiber},
  {Rudnick}, {Barden}, {Franx}, {Rix}, {Caldwell}, {McIntosh}, {Toft},
  {H{\"a}ussler}, {Zirm}, {van Dokkum}, {Labb{\'e}}, {Moorwood},
  {R{\"o}ttgering}, {van der Wel}, {van der Werf}, \& {van
  Starkenburg}}]{2006Trujillo}
{Trujillo}, I., {F{\"o}rster Schreiber}, N.~M., {Rudnick}, G., {Barden}, M.,
  {Franx}, M., {Rix}, H., {Caldwell}, J.~A.~R., {McIntosh}, D.~H., {Toft}, S.,
  {H{\"a}ussler}, B., {Zirm}, A., {van Dokkum}, P.~G., {Labb{\'e}}, I.,
  {Moorwood}, A., {R{\"o}ttgering}, H., {van der Wel}, A., {van der Werf}, P.,
  \& {van Starkenburg}, L. 2006, \apj, 650, 18

\bibitem[{{Turk} {et~al.}(2011){Turk}, {Smith}, {Oishi}, {Skory}, {Skillman},
  {Abel}, \& {Norman}}]{turk11}
{Turk}, M.~J., {Smith}, B.~D., {Oishi}, J.~S., {Skory}, S., {Skillman}, S.~W.,
  {Abel}, T., \& {Norman}, M.~L. 2011, \apjs, 192, 9

\bibitem[{{Weingartner} \& {Draine}(2001)}]{weingartner01}
{Weingartner}, J.~C., \& {Draine}, B.~T. 2001, \apj, 548, 296

\bibitem[{{Wilkins} {et~al.}(2011){Wilkins}, {Bunker}, {Stanway}, {Lorenzoni},
  \& {Caruana}}]{2011Wilkins}
{Wilkins}, S.~M., {Bunker}, A.~J., {Stanway}, E., {Lorenzoni}, S., \&
  {Caruana}, J. 2011, \mnras, 417, 717

\bibitem[{{Wise} \& {Cen}(2009)}]{2009Wise}
{Wise}, J.~H., \& {Cen}, R. 2009, \apj, 693, 984

\bibitem[{{Yajima} {et~al.}(2011){Yajima}, {Choi}, \& {Nagamine}}]{2011Yajima}
{Yajima}, H., {Choi}, J.-H., \& {Nagamine}, K. 2011, \mnras, 412, 411

\bibitem[{{Yan} {et~al.}(2010){Yan}, {Yan}, {Zamojski}, {Windhorst},
  {McCarthy}, {Fan}, {R{\"o}ttgering}, {Koekemoer}, {Robertson}, {Dav{\'e}}, \&
  {Cai}}]{2010Yan}
{Yan}, H., {Yan}, L., {Zamojski}, M.~A., {Windhorst}, R.~A., {McCarthy}, P.~J.,
  {Fan}, X., {R{\"o}ttgering}, H.~J.~A., {Koekemoer}, A.~M., {Robertson},
  B.~E., {Dav{\'e}}, R., \& {Cai}, Z. 2010, ArXiv e-prints

\end{thebibliography}

\end{document}